\documentclass[12pt]{iopart}
\usepackage{iopams}
\usepackage[english]{babel} 

\newcommand{\spsp}{S_{\rm psp}}

\newcommand{\scan}{S_{\rm can}}
\newcommand{\smc}{S_{\rm mc}}
\newcommand{\ope}[1]{\widehat{#1}} 
\newcommand{\tihop}{\ope{\rho}}
\newcommand{\A}{\ope{A}} 
\newcommand{\nham}{\ope{h}}
\newcommand{\hameff}{\ope{H}_{\rm eff}}
\newcommand{\escm}{\ope{\omega}} 
\newcommand{\ham}{\ope{H}}
\newcommand{\pref}{\ope{\Phi}} 

\newcommand{\qtihop}{\tihop^{\,q}} 
\newcommand{\vep}{\varepsilon}
\newcommand{\ket}[1]{|#1\rangle} 
\newcommand{\bra}[1]{\langle #1|}
\newcommand{\mean}[1]{\langle #1\rangle} 
\newcommand{\ci}{\rmi}
\newcommand{\id}{{\rm id}}
\newcommand{\re}{{\rm Re\,}} 
\newcommand{\im}{{\rm Im\,}}
\newcommand{\half}{\case12} 
\newcommand{\qand}{\quad{\rm and}\quad} 
 \newcommand{\text}[1]{{\rm #1}}
\newcommand{\order}[1]{\Or{(#1)}} 
\newcommand{\R}{{\mathbb R}}

\newtheorem{theorem}{Theorem} 
\newtheorem{corollary}[theorem]{Corollary} 

\newtheorem{axiom}{Axiom} 
\newtheorem{axiomp}{Axiom}

\begin{document} 

\jl{1}

\title{On the use of non-canonical quantum statistics}

\author{Jani Lukkarinen\footnote[1]{E-mail address: %
 {\tt jani.lukkarinen@helsinki.fi} }} 
 \address{Helsinki Institute of Physics, PO Box 9, %
 00014 University of Helsinki, Finland}

\begin{abstract}
We develop a method using  a coarse graining of the energy
fluctuations of an equilibrium quantum system which produces simple
parametrizations for the behaviour of the system.  As an application,
we use these methods to gain more understanding
on the recently developed Tsallis statistics and we show how
energy reparametrizations can be used for producing
other similar generalizations of the Boltzmann-Gibbs statistics.
We conclude on a brief discussion on the role of entropy and
the maximum entropy principle in thermodynamics and in quantum 
statics.
\end{abstract}

\pacs{05.30.Ch, 02.50.Kd, 03.65.Db}

\submitted
\maketitle

\section{Introduction}

The usefulness of the canonical ensemble in statistical mechanics is
remarkable.  The standard explanation for this success relies on 
taking the
thermodynamical limit which corresponds to increasing the volume of
the system to infinity while keeping all the relevant intensive
quantities, i.e.\ densities, fixed and finite.  From this point of
view, the canonical ensemble should not have much utility for small
systems consisting of only a few particles.  This is not completely
true and canonical expectation values can be used also for small
systems \cite{jml:gauss},  although not very accurately in the direct
manner used in thermodynamics.

We shall try to generalize the ideas leading to the standard
thermodynamics so that these methods could be applied in the analysis
of small systems.  We shall first concentrate on finding practical
approximations which capture the large scale behaviour of an
equilibrium system.  For this purpose, we present in sections
\ref{sec:vancan} and \ref{sec:PSPA} a coarse graining
procedure which justifies a modified Gaussian ensemble and we derive
a positive saddle point (PSP) approximation which can be
applied for further simplifying the Gaussian results.

As an application of these methods, we present in section 
\ref{sec:essmc} how the
PSP-approximation leads to a generalization of thermodynamics  
for essentially isolated systems. As a second application, we derive 
the recently proposed non-extensive Tsallis statistics \cite{tsa88}
as a PSP-approximation.  We conclude on a
discussion of implications to Tsallis thermodynamics and 
to the maximum entropy principle in the last two sections,
\ref{sec:tsaexpl} and \ref{sec:maxS}.

\section{Preliminaries in quantum statistics}

The standard approach to quantum statistics \cite{Balian1} is  defined
by using a density matrix $\tihop$, which is a non-negative,
hermitian, trace-class operator normalized to one and which gives the
expectation value of an observable $\A$ by the formula
$\mean{\A}=\Tr(\A\tihop)$.  In some complete eigenbasis
$\ket{\psi_i}$, the density matrix can thus be expanded as
\begin{equation}\label{e:defrho}
\tihop=\sum_i p_i \ket{\psi_i}\bra{\psi_i},
\end{equation}
where the eigenvalues $p_i$ are non-negative, $p_i\ge 0$, and they
satisfy the normalization condition $\sum_i p_i = 1$.

Suppose next that the system has a discrete energy spectrum with
finite degeneracies, which in quantum mechanics is achieved for every
sufficiently binding potential. An equilibrium, i.e.\
time-independent,  ensemble is then given by a density matrix which
has time-independent eigenvalues and which satisfies the
Liouville--von Neumann equation $[\ham,\tihop]=0$. In this case, the
eigenvectors $\psi_i$ can be chosen so that they are also energy
eigenvectors; let $E_i$ denote the eigenvalue corresponding to
$\psi_i$.

Assume also that energy is the only relevant conserved quantity in
this equilibrium system. By this we mean that the probability of
finding any energy eigenstate depends only on energy and not on any
other conserved quantum numbers, i.e.\ we require that $p_i=p_j$ for
every pair of indices with $E_i=E_j$.  Then we can find a function $f$
such that $\tihop = f(\ham)$, which is just another notation for
demanding that $p_i = f(E_i)$ in (\ref{e:defrho}).   It is also
obvious that in this case $f$ can be chosen as a smooth, positive
function such that $\int\!\rmd E\, f(E)< \infty$.  Then we can define
a smooth probability density $F$ by the formula  
$F(E) = f(E)/\int\! f$ using which the density operator becomes
\begin{equation}\label{e:defF}
\tihop = \frac{F(\ham)}{\Tr F(\ham)}.
\end{equation}

We shall call any smooth probability density $F$ which satisfies
(\ref{e:defF}) a {\em fluctuation spectrum}\/ of the system.  If we
require that the energy spectrum is bounded from below, we can
normalize the Hamiltonian so that the lowest  eigenvalue is strictly
positive and then choose  a fluctuation spectrum with a support on the
positive real axis.

Finally, let us define a few mathematical tools needed in the
following discussion.  We shall denote the Gaussian probability
density with mean $0$ and standard deviation $\lambda$ by 
$G_\lambda$, i.e.\
\[
G_\lambda(x) = \frac{1}{\sqrt{2 \pi \lambda^2}} 
\exp\!\left(-\frac{x^2}{2\lambda^2} \right).
\]
For a pointwise application of inverse Fourier-transforms, we shall
also need the class $\mathcal{S}$ of Schwarz test functions, 
also called rapidly decreasing functions.  They are infinitely many
times differentiable, i.e.\ smooth, functions which have the property
that the function and any of its derivatives vanish faster than any
power at infinity---the distribution $G_\lambda$ is a prime example of
a rapidly decreasing function.  For a more precise definition 
see e.g.\ chapter~6 of the book on functional analysis by
Rudin~\cite{rudin:fa}.

\section{An overview of the results for 
  standard ensembles}\label{sec:vancan}

The purpose of this section is to give a brief review of the main
results in the special case when they lead to the
usual canonical and Gaussian ensembles.  We have chosen this 
peculiar order of presentation, so that there would be a concrete
application which can be kept in mind when going through the more
abstract results in the following section.  It also serves as an
introduction to the methods we apply in the following section.
We do not imply that these results are obvious---we shall go
through their derivations in detail in section \ref{sec:PSPA}.

Let $F$ be a fluctuation spectrum of an
equilibrium system for which energy is the only relevant 
conserved quantity and let $\A$ be a positive observable.
Assume also that
\newcounter{alp} \renewcommand{\thealp}{A\arabic{alp}}
\begin{list}{\thealp.}{\usecounter{alp}}
\item\label{it:trace} The system is canonical: its spectrum is bounded
from below and  $\Tr \rme^{-\beta\ham} <\infty$ for all $\beta>0$.
\item\label{it:expdecay} The energy fluctuations decay faster than
exponentially at high energies: $\rme^{\beta E} F(E)$ is a rapidly
decreasing function for all real $\beta$. 
\item\label{it:canobs} The observable is canonical:
$\Tr(\A \rme^{-\beta\ham})<\infty$ for all $\beta >0$.
\end{list}
We shall call a fluctuation spectrum satisfying the condition
\ref{it:expdecay} precanonical and we shall use the name precanonical
system for those systems which are canonical and have precanonical
fluctuations.  Note also that by \ref{it:trace} we can choose $F$
so that its support is bounded from below and thus
\ref{it:expdecay} is a restriction for the behaviour of the
fluctuations only near $E\to+\infty$.

The assumptions \ref{it:trace}--\ref{it:canobs} allow for all
$\beta>0$ the integral representation
\begin{equation}\label{e:trAcan}
\Tr\!\left(\A F(\ham)\right) = 
 \int_{\beta-\ci \infty}^{\beta+\ci \infty} 
 \frac{\rmd w}{ 2\pi\ci} \bar{F}(w)  
 \Tr\!\left(\A\rme^{-w \ham}\right),
\end{equation}
where $\bar{F}$ is the Laplace transform of $F$  and the integrand in
the above equation is an analytic function in the half-plane $\re
w>0$.  We can use saddle point methods for finding a better
integration contour and approximations to the integral. In fact,
usually there exists a unique positive
saddle point $\beta_A$ which will be the best choice for $\beta$ in
(\ref{e:trAcan}) as  the integration contour will then go through this
saddle point via the path of steepest descent.  

The saddle point approximation around $\beta_A$ will 
then yield ``the best possible canonical approximation'' of the trace.
This saddle point approximation can be expressed in terms 
of $\bar{g}_A(\beta)=\mean{\A\ham}/\mean{\A}$ 
and
$\sigma^2_A(\beta)=\mean{\A(\ham-\bar{g}_A)^2}/\mean{\A}$,
where all expectation values are taken in the canonical ensemble at
inverse temperature $\beta$, and in terms of
the following two characteristics of the fluctuation spectrum,
\begin{equation}\label{e:defab}
a(\beta)  = \frac{\int\! \rmd x F(x) \rme^{\beta x} x  
        }{ \int\! \rmd x F(x) \rme^{\beta x} } \qand
b^2(\beta) = \frac{\int\!\rmd x F(x)\rme^{\beta x} [x-a(\beta)]^2
    }{\int\! \rmd x F(x) \rme^{\beta x}}.
\end{equation}
The saddle point equation is then 
\[
a(\beta) = \bar{g}_A(\beta),
\]
and there is unique a positive solution, $\beta_A$,
to this equation provided 
\begin{equation}\label{e:ispsp}
a(0) < \lim_{\beta\to 0^+} \bar{g}_A(\beta).
\end{equation}
When this is the case, 
the saddle point approximation around $\beta_A$ yields
\begin{equation}\label{e:cang1}
\Tr\!\left(\A F(\ham)\right) \approx
 \left[2\pi (b^2 + \sigma^2_A) \right]^{-\half}  
 \rme^{a\beta_A+\half b^2\beta_A^2} 
 \Tr\!\left(\A\rme^{-\beta_A \ham}\right),
\end{equation}
where all functions are to be evaluated at $\beta=\beta_A$.

Suppose now that the positive saddle point approximation is good for
the partition function, i.e.\ for $\A=\ope{1}$, and let $\beta_0$ be
the corresponding saddle point value.  Assume also that 
$\A$ is an observable which does not alter 
the canonical distribution much,
$\bar{g}_A(\beta) = \mean{\A\ham}_\beta/\mean{\A}_\beta 
\approx \mean{\ham}_\beta$ and 
$\sigma_A^2 \approx \mean{\ham^2} - \mean{\ham}^2$.
Then we can use $\beta_0$ also for the saddle
point approximation of $\Tr[\A F(\ham)]$ and arrive at the usual
canonical formula
\[
 \frac{\Tr\!\left[\A F(\ham)\right]}{\Tr F(\ham)} \approx   
 \frac{\Tr\!\left[\A\rme^{-\beta_0 \ham}\right]}{
    \Tr \rme^{-\beta_0 \ham}}.
\]

For large systems, we are typically only interested in
``total'' observables like total energy and total particle
number.  These observables depend only on
the behaviour of the system at the scale of its total energy and,
therefore, we would  expect to need only information about the
behaviour of the system coarse grained up to a scale $\Lambda$ which
is less than the total energy but larger than the microscopic energy
scales.

For precanonical systems, it is
possible to define a Gaussian approximation which will yield
accurate results for most large scale\footnote{These can be defined as
  observables whose expectation value 
  changes less than some preset limit when the fluctuation spectrum is
  coarse grained.} 
canonical observables:  if we smooth the fluctuation
spectrum by taking a convolution with a Gaussian distribution 
with variance $\Lambda$, then after defining $L=a(0)$ and 
$\lambda^2=b^2(0)$,
we get in the limit $\Lambda\to\infty$,
\begin{equation}\label{e:standardg}
 \Tr\!\left[\A\, (F\ast G_\Lambda)(\ham)\right] = 
  \Tr\!\left[\A\, (G_\lambda\ast G_\Lambda)(L-\ham)\right] 
  \left( 1 + \order{\Lambda^{-3}} \right).
\end{equation}
This choice of parameters $L$ and $\lambda^2$ 
is natural since they are the
expectation value and variance of the probability distribution $F$,
as can be seen from (\ref{e:defab}).  They are, in fact, also the best
possible, since in theorem \ref{th:gaussconv} we shall prove that any
other choice of $L$ and $\lambda^2$ 
guarantees only a slower convergence in $1/\Lambda$.  

The precise conditions for an application of the theorem
are discussed in section \ref{sec:gaussapp} and
one set of sufficient conditions is given by
\begin{enumerate}
\item\label{it:hbarinf} $\lim_{\beta\to 0^+} \bar{g}_A(\beta) = \infty$,
\item\label{it:bndness} $\beta\bar{g}_A(\beta)$ and 
$\beta\sigma_A^2(\beta)/\bar{g}_A(\beta)$ stay bounded in the limit
$\beta\to 0^+$.
\end{enumerate}
It is relatively easy to check that these conditions hold, 
for example, for all energy moments ($\A=\ham^k$)
when the system consists of $N$ particles in a harmonic potential. 
In addition, by (\ref{e:ispsp}), condition 
(\ref{it:hbarinf}) also guarantees the existence of a positive saddle
point solution and thus of a canonical approximation.

If these two conditions hold also for $\A=\ope{1}$, we 
can then see that the Gaussian ensemble yields an accurate
approximation to the original ensemble
\[
 \frac{\Tr\!\left[\A F(\ham)\right]}{\Tr F(\ham)}
 \approx   
\frac{\Tr\!\left[\A G_\lambda(L-\ham)\right] }{ 
  \Tr G_\lambda(L-\ham) },
\]
at least for all sufficiently large scale observables $\A$.  

\section{The positive saddle point (PSP) approximation}\label{sec:PSPA} 

In the preceding section we anticipated the derivation of the 
canonical ensemble from a saddle point approximation under very general
assumptions.  Is there any need for non-canonical ensembles?  In
principle, the canonical approximation can fail in several different
ways as the following examples illustrate.
\begin{enumerate}
\item\label{it:noncan} The density of states increases faster than
exponentially or is not discrete and thus  the canonical ensemble is
ill-defined.
\item\label{it:polfluc} $F$ decays only e.g.\ polynomially in energy.
Then the above arguments relying on analyticity will not hold and the
Gaussian approximation might not be applicable.
\item\label{it:noncanobs} The addition of the observable
$\A$ alters the canonical energy distribution too radically and 
either we cannot use $\beta_A=\beta_0$ or, even worse, 
$\Tr(\A \rme^{-\beta\ham})=\infty$.
\item\label{it:usegaus} Canonical ensemble gives too rough an
approximation for the real statistics.  This will always
eventually happen when observables resolving ever smaller energy 
scales are inspected, unless the original fluctuation spectrum is 
{\em exactly}\/ canonical.
\end{enumerate}

Let us now first concentrate on solving the ``exotic'' part of the
problem, the exponential increase of density of states
(cases \ref{it:noncan} and \ref{it:noncanobs}) and the 
possibility of non-exponentially decaying fluctuations 
(case \ref{it:polfluc}).  
A solution in both cases is to re\-parametrize 
the energy so that in the
new variables the asymptotic high-\-energy behaviour is 
sufficiently regular.

Let $g(E)$ be the function which performs this reparametrization, 
i.e.\ let $g$ is be an orientation preserving
diffeomorphism from the real line onto
itself.   Since we do not want to restrict the speed of the
growth of the
density of states in any way, we do not assume anything more of the
diffeomorphism at this stage, only that it regularizes the
high-energy behaviour of both the density of states and the
fluctuation spectrum: 
we assume that in the reparametrized variables the density of states
(with the effect of the observable included) does not increase
exponentially and that the fluctuation spectrum decreases at least
exponentially.  

Expressed mathematically, we assume that there exists real numbers
$\beta_-$ and $\beta_+$ such that $\beta_-\le 0 < \beta_+$ and  
\renewcommand{\thealp}{B\arabic{alp}}
\begin{list}{\thealp.}{\usecounter{alp}}
\item\label{it:gobs}
For all $\beta>\beta_-$,
$\A$ is a positive $g$-bounded observable:
$\Tr\bigl(\A \rme^{-\beta g(\ham)}\bigr) <\infty$.
\item\label{it:gexpdecay} The reparametrized  fluctuations decrease
at least exponentially at high energies: $\rme^{\beta x} F(g^{-1}(x))$
is a rapidly decreasing function for all $\beta<\beta_+$.
\end{list}
We also assume that these parameters have
been chosen in the best possible manner,
\begin{eqnarray}
\beta_-  = \inf\,\{\beta\,|\,
  \Tr\bigl(\A\rme^{-\beta {g(\ham)}}\bigr) <\infty\},\\ 
\beta_+  = \sup\,\{\beta\,|\, 
 \rme^{\beta x} F(g^{-1}(x))\in \mathcal{S}\},
\end{eqnarray}
and thus either of them might be infinite.  Note also that $\beta_-$
depends on the choice of the observable $\A$.

It looks like we lost some generality at this stage.  However, by
adding a suitable part of the energy dependence 
to the observable, i.e.\ by
replacing $\A$ by $\A\Phi(\ham)$ where $\Phi(E)$ is a 
suitable function, it is usually possible to find
such a reparametrization
$g$ that $\beta_-$ and $\beta_+$ satisfy the
requirement $\beta_-\le 0 < \beta_+$.  For example, if the density of
states increases like $\rme^{\beta'\ham}$ choosing
$\Phi(E)=\rme^{-\beta'E}$ would allow still using $g=\id$.  

Suppose now that we would like to approximate the behaviour of the
system by using the reparametrized Gaussian ensemble, 
$\tihop = G_\lambda(L-g(\ham))$, 
or the reparametrized canonical ensemble,
$\tihop = \rme^{-\beta g(\ham)}$.  It would be natural to assume
that the best approximation of this kind follows if we choose 
the parameters $L$ and $\lambda$ as the reparametrized expectation
value and variance of $F$ and $\beta$ from the saddle point
approximation of the trace in the partition function.  We will
now concentrate on deriving conditions under which this
intuition is valid and on finding estimates 
for the errors induced by the approximations. 

\subsection{Canonical approximation of a trace}\label{sec:canaptr}

It follows easily from assumption \ref{it:gexpdecay} that the function 
\begin{equation}\label{e:Fbardef}
\bar{F}(w) = \int_{-\infty}^\infty\!\rmd x\, \rme^{x w} F(g^{-1}(x))
\end{equation}
is well-defined and analytic in the region $\re w<\beta_+$.  Since $g$
was a diffeomorfism and we can typically choose $F$ with a support on
the positive real axis, we can usually
also define this ``$g$-transform'' of $F$ by
\[
\bar{F}(w) = \int_0^\infty\!\! \rmd E g'(E) \rme^{g(E) w} F(E).
\]
When $g(E)=E$ this is just the Laplace transform  and if $g(E)=\ln E$
it coincides with the Mellin transform.

As the inverse Fourier-transformation can be taken pointwise for 
rapidly decreasing functions, we have pointwise 
for all $\beta<\beta_+$,
\begin{equation}\label{e:FF}
F(E) = \int \frac{\rmd p}{ 2\pi} \rme^{\ci p g(E)-\beta g(E)}
  \bar{F}(\beta-\ci p) = \int_{\beta-\ci \infty}^{\beta+\ci \infty}
  \frac{\rmd w}{ 2\pi\ci} \bar{F}(w) \rme^{-w g(E)}.
\end{equation}
By applying the representation (\ref{e:FF}) we get then for any
observable satisfying \ref{it:gobs} an integral representation valid
for $\beta_-<\beta<\beta_+$,
\begin{equation}\label{e:trF}
\Tr\bigl(\A F(\ham)\bigr) = \int_{\beta-\ci \infty}^{\beta+\ci \infty}
\frac{\rmd w}{ 2\pi\ci} \bar{F}(w) \Tr\bigl(\A \rme^{-w g(\ham)}\bigr).
\end{equation}
The boundedness of the trace also implies that the function
$\Tr\bigl(\A \rme^{-w g(\ham)}\bigr)$ is analytic in the half plane
$\re w>\beta_-$ and, therefore, the integrand in (\ref{e:trF}) is an
analytic function in the strip $\beta_-<\re w<\beta_+$ and we proceed
to consider its evaluation by the method of steepest descent.

For writing down the saddle point approximation it will be useful to
define
\begin{equation}\label{e:defpars}
\fl\eqalign{
a(w) = \frac{\int\! \rmd x F(g^{-1}(x)) \rme^{w x} x  
        }{ \int\! \rmd x F(g^{-1}(x)) \rme^{w x} } \qand
b^2(w) = \frac{\int\!\rmd x F(g^{-1}(x))\rme^{w x} [x-a(w)]^2
    }{\int\! \rmd x F(g^{-1}(x)) \rme^{w x}}, \\ 
\bar{g}(w) = \frac{ \Tr\left(\A \rme^{-w g(\ham)} g(\ham) \right)  
    }{\Tr\biggl(\A\rme^{-w g(\ham)}\biggr)} \qand
\sigma^2(w) = \frac{\Tr\biggl(\A \rme^{-w g(\ham)}  
  \bigl[g(\smash{\ham})-\bar{g}(w)\bigr]^2 \biggr)
    }{\Tr\biggl(\A\rme^{-w g(\ham)}\biggr)}, 
}
\end{equation}
which are well-defined in the whole strip apart from the 
countable set of zeros of the denominators.

In terms of these quantities, the saddle point equation becomes
\begin{equation}\label{e:gsp}
a(w) - \bar{g}(w) = 0,
\end{equation}
and the second derivative of the logarithm of the integrand is given by
\begin{equation}\label{e:g2nd}
b^2(w) + \sigma^2(w).
\end{equation}
For real values of $w$, both $b^2$ and $\sigma^2$ are strictly
positive and thus the second derivative becomes a strictly positive
quantity on the real part of the strip.  Therefore, the first
derivative defining the saddle point equation is a strictly increasing
function on the interval $\beta_-<w<\beta_+$ and there exists a real
saddle point on this interval if and only if
\begin{equation}\label{e:isgsp}
a(\beta_-) < \bar{g}(\beta_-)\qand a(\beta_+) > \bar{g}(\beta_+).
\end{equation}
If the real saddle point exists, it is obviously unique and choosing
$\beta$ in (\ref{e:trF}) equal to this saddle point value will lead
to an integration contour which goes through the real saddle point via
the steepest descent path.

Assume then that the real saddle point exists, 
denote it by $\beta_A$,
and parametrize the integration variable in (\ref{e:trF})
as $w=\beta_A+\ci\alpha$.
If the integrand, as a function of $\alpha$, is strongly
concentrated in such a neighbourhood of the origin that the 
quadratic approximation of the logarithm of the 
integrand is admissible, then the saddle
point approximation yields
\[
\Tr\bigl(\A F_\Lambda(\ham)\bigr) \approx 
 \frac{1}{\sqrt{2\pi(b^2(\beta_A)+\sigma^2(\beta_A))}}
 \bar{F}(\beta_A) \Tr\bigl(\A \rme^{-\beta_A g(\ham)}\bigr).
\]
Note that this can happen, since the maximum of the absolute 
value of the integrand is at $\alpha=0$, 
although it cannot be guaranteed without further
assumptions.  For a derivation of bounds for the accuracy of the
saddle point approximation, we shall need to make a detour via 
a Gaussian approximation. 

\subsection{Gaussian approximation of a trace}\label{sec:gaussapp}

Suppose we want to coarse grain the fluctuation spectrum like in the
previous section and that way deduce what kind of Gaussian 
approximation would be most accurate for large scale observables.  
The use of a Gaussian convolution is not justified now
since we do not know if  the resulting function will give a
trace-class operator.  Moreover, the transformed function might not
satisfy the decay condition  \ref{it:gexpdecay} any more and we could
not use the above integral representation for it.

There is, however, a natural generalization of the usual coarse
graining procedure which will give functions satisfying condition
\ref{it:gexpdecay}: we shall  perform the coarse graining by a
convolution with a Gaussian distribution  
{\em in the reparametrized space\/},
\[
\fl
F_\Lambda(E) = \int\!\! \rmd y F(g^{-1}(y))G_\Lambda(g(E)-y) 
  = \int\!\! \rmd x F(x)  g'(x) \frac{1}{\sqrt{2 \pi \Lambda^2}}
 \rme^{-\frac{1}{ 2 \Lambda^2} (g(E)-g(x))^2}.
\]
It is obvious that this transformation satisfies the following
semigroup-property,
\[
(F_\Lambda)_{\Lambda'} = F_{\sqrt{\Lambda^2+{\Lambda'}^2}},
\]
and that the $g$-transform changes simply by multiplication,
\[
\bar{F}_\Lambda(w) = \rme^{\half\Lambda^2 w^2} \bar{F}(w).
\]

As the Gaussian distribution is a rapidly decreasing function,  its
convolution leaves the class $\mathcal{S}$ invariant.  Therefore, if
$\rme^{\beta x} F(g^{-1}(x))$ is rapidly decreasing, so is
$\rme^{\beta x} F_\Lambda(g^{-1}(x))$,  since clearly
\[
\rme^{\beta x} F_\Lambda(g^{-1}(x)) = \rme^{\half\beta^2\Lambda^2}
 \int\!\rmd y \,\rme^{\beta y} F(g^{-1}(y))
  G_\Lambda(x-\beta\Lambda^2-y).
\]
This proves that if $F$ satisfies condition \ref{it:gexpdecay} 
for some $\beta_+$ then so does $F_\Lambda$.

By the above results, we now have an integral representation
\begin{equation}\label{e:trFL} 
\Tr\bigl(\A F_\Lambda(\ham)\bigr)  = 
 \int_{\beta-\ci\infty}^{\beta+\ci \infty}  
  \frac{\rmd w}{ 2\pi\ci} \rme^{\half\Lambda^2 w^2}  
  \bar{F}(w) \Tr\bigl(\A \rme^{-w g(\ham)}\bigr),
\end{equation}
which is valid for all $\beta_-<\beta<\beta_+$ and $\Lambda\ge 0$.
We would now like to make a quadratic approximation of $\bar{F}(w)$
which would then lead to a Gaussian trace.  In fact, this is often
possible in the limit of large $\Lambda$, but since the discussion
gets a bit technical at this point, we leave the proof of this 
to \ref{sec:gausspf} and state only the results here.

Define the functions $a$, $b^2$, $\bar{g}$ and $\sigma^2$ by equations
(\ref{e:defpars}).
Then $a(0)$ and $b^2(0)$ are the expectation value and the
variance of the reparametrized fluctuation spectrum and we proceed to
show that these values yield the best possible approximation of a
Gaussian form.  If $\beta_-=0$, we also need to assume that
the following three conditions are satisfied:
\renewcommand{\thealp}{G\arabic{alp}}
\begin{list}{\thealp.}{\usecounter{alp}}
\item\label{Git:first}\label{Git:issp} $a(0) < \bar{g}(0)$ 
\item\label{Git:gbarcond} There exists a constant $c\ge 0$ 
for which
\begin{equation}\label{e:gbarcond}
\beta^{1+2 c}\bar{g}(\beta)
 \mbox{ stays bounded in the limit }\beta\to 0^+.
\end{equation}
\item\label{Git:last}\label{Git:sigmacond} 
$\beta\sigma^2(\beta)/\bar{g}(\beta)$ stays
bounded in the limit $\beta\to 0^+$.
\end{list}

With these notations, the following theorem, whose proof we have
included in \ref{sec:gausspf}, holds:
\begin{theorem}\label{th:gaussconv}
Let $L=a(0)$ and $\lambda^2=b^2(0)$ and assume that \ref{it:gobs} and
\ref{it:gexpdecay} are satisfied for some $\beta_-\le 0 < \beta_+$.  
If $\beta_- = 0$ and
\ref{Git:first}--\ref{Git:last} hold, 
then in the large scale limit
$\Lambda\to\infty$ for all $c\ge 0$ satisfying (\ref{e:gbarcond}),
\begin{equation}\label{e:gtrapp}
 \Tr\bigl(\A F_\Lambda(\ham)\bigr) = 
  \bar{F}(0) \Tr\left[\A G_{\sqrt{\lambda^2+\Lambda^2}}(L-g(\ham))\right] 
  \left( 1 + \order{\Lambda^{-\frac{3}{1+c}}} \right),
\end{equation}
and if $\beta_-<0$, then (\ref{e:gtrapp}) is true for $c=0$.

In addition, for any other choice of $L$ or $\lambda^2$ only a slower
convergence in this limit can be guaranteed. 
\end{theorem}
Since $G_{\sqrt{\lambda^2+\Lambda^2}}=(G_\lambda)_{\Lambda}$, 
we can thus conclude
that, when the quality of the approximation is measured by expectation
values of large scale observables,
the best approximation of the Gaussian form for $F(\ham)$ is given by 
\begin{equation}\label{e:gaussapp}
F(\ham) \approx \bar{F}(0) G_\lambda(L-g(\ham)).
\end{equation}
Therefore, we have proved that
having either $\beta_-<0$ or conditions
\ref{Git:first}--\ref{Git:last} satisfied will lead to the
intuitive approximation mentioned earlier in section \ref{sec:canaptr}.

The standard situation we examined in section \ref{sec:vancan} implies
that $\beta_-=0$, $\beta_+=\infty$ and $g(x)=x$.  Thus we then need to
satisfy the latter three conditions.  In fact, it is straightforward to
see that the conditions (\ref{it:hbarinf}) and (\ref{it:bndness}) 
given in section \ref{sec:vancan}
do imply that \ref{Git:issp} is satisfied for any $a(0)$, 
that we can choose $c=0$ in condition 
\ref{Git:gbarcond}, and that
\ref{Git:sigmacond} holds.  
Theorem \ref{th:gaussconv} will then justify equation (\ref{e:standardg}).

\subsection{Accuracy of the canonical approximation}\label{sec:canacc}

The canonical approximation was derived in section \ref{sec:canaptr}
from a real saddle point approximation of the trace.  Estimating
the accuracy of this approximation is, however, quite difficult.  We
shall in this section present estimates for the accuracy of the saddle
point approximation in the special case when the fluctuation spectrum
is Gaussian.  By the results of the previous section, this can then be
used also for more general fluctuation spectra, 
if the observable is of
sufficiently large scale and the conditions for the use of the Gaussian
approximation are valid.  

Assume thus that the fluctuation spectrum is
$F(\ham)=G_\lambda(L-\ham)$, when $\beta_+=\infty$,
and let $\beta$ be some value greater than $\beta_-$.  
We shall later see that the best bounds are
obtained if $\beta$ is the solution to the saddle point equation
\begin{equation}\label{e:ggsp}
\beta\lambda^2 + L - \bar{g}(\beta) = 0.
\end{equation}
However, since we shall also need estimates for how far from the saddle
point value $\beta$ can be chosen before the saddle point
approximation loses its accuracy, we shall not fix the value of
$\beta$ yet.

The bounds will be expressed in terms of the ratio
$R=\lambda/\sigma(\beta)$.
Since the derivations of the bounds are again quite technical, we
postpone them into \ref{sec:canaccder} and present only
the results here.  The real saddle point approximation to the Gaussian
trace is there shown to yield 
\begin{eqnarray}
\fl\Tr\!\left[\A G_{\lambda}(L-g(\ham))\right] 
\approx
  \rme^{\beta L +\half \beta^2 \lambda^2 }  
 G_{\sqrt{\lambda^2+\smash{\sigma^2(\beta)}}}
    (L+\beta\lambda^2-\bar{g}(\beta))
 \Tr\!\left[\A\rme^{-\beta g(\ham)}\right], \label{e:PSPA}
\end{eqnarray}
and we call the right hand side ``saddle point approximation'' and
denote it by ``PSPA'' even
when $\beta$ does not satisfy the saddle point equation (\ref{e:ggsp}).

The first bound for the accuracy of
this approximation is obtained by using Jensen's
inequality very analogously to what was done in \cite{jml:gauss}.
This shows that for any $\beta$ the ratio of the Gaussian
trace and its PSP-approximation is bounded by
\begin{eqnarray}
\fl \exp\biggl[\frac{
  \left( L+\beta\lambda^2-\bar{g}(\beta) \right)^2
 }{2(\lambda^2+\sigma^2(\beta))} \biggr]
\biggl(1+\frac{1}{R^2}\biggr)^{\half} \nonumber \\
\lo\ge 
  \frac{ \Tr\bigl[\A G_{\lambda}(L-g(\ham))\bigr]
  }{{\rm PSPA}}\ge \biggl(1+\frac{1}{R^2}\biggr)^{\half}
  \exp\biggl[-\frac{1}{2 R^2}\biggr]. \label{e:canJB}
\end{eqnarray}
These inequalities prove that, if $\beta$ is close to the saddle point
value and $R\gg 1$, the relative error from the PSPA
becomes negligible.

The second bound for the relative error is given for 
$\Delta_{\rm PSPA}$ in 
\[
  \frac{ \Tr\bigl[\A G_{\lambda}(L-g(\ham))\bigr]
  }{{\rm PSPA}} = 1 + \Delta_{\rm PSPA}
\]
and it depends on the behaviour of $\sigma^2(\beta+\ci \alpha)$ near
$\alpha=0$.   Since $\sigma^2(\beta)\ne 0$, we can for all $0<r<1$
find a $\rho_r>0$ such that
\begin{equation}\label{e:defcanrc}
\left|\frac{\sigma^2(\beta+\ci\alpha)}{\sigma^2(\beta)}-1\right| 
 \le r \text{,\ for\ all\ } -\rho_r\le\alpha\le \rho_r.
\end{equation}
The bound is then expressed in terms of $r$ and $\rho_r$ as
\begin{eqnarray}\label{e:Dcanbound}
\fl \exp\biggl[\frac{
   \left( L+\beta\lambda^2-\bar{g}(\beta) \right)^2
   }{2(\lambda^2+\sigma^2(\beta))} \biggr]
 |\Delta_{\rm PSPA}|  \nonumber \\ 
 \lo\le \frac{3}{2} \frac{r}{1+ R^2} 
  \smash{\biggl(1- \frac{r}{1+R^2}\biggr)^{-\case32}} 
  + \rme^{-\smash{\half} \rho_r^2 \lambda^2} \biggl[ 1 + 
  \biggl( 1 + \frac{1}{R^2}\biggr)^{\half} \biggr].
\end{eqnarray}
Clearly, the bound is informative only if $\lambda$ is so large that
$\lambda\rho_r\ge 1$.  

If $\beta$ is so close to the saddle point value that 
$\left(L+\beta\lambda^2-\bar{g}(\beta)\right)^2 
\ll \sigma^2(\beta)+\lambda^2$ and if $r$
can be chosen so that $|\Delta_{\rm PSPA}|$ is small, then 
the PSP-approximation of a Gaussian trace is very accurate.  
By applying the first bound (\ref{e:canJB}), we can see
that this will happen after a coarse graining with
a large enough $\Lambda$, if $\beta$ can be chosen so that
$\Lambda\gg\sigma(\beta)$.
Unfortunately, unlike was erroneously claimed in section 4 in
\cite{jml:gauss}, the last condition can not be satisfied in a typical
thermodynamical limit and the more complicated second bound has to be
applied in these cases.

\subsection{Expectation values}\label{sec:evs}

So far we have inspected the approximation of one trace only.  The
statistical expectation values are ratios of two traces  and
as some of the terms cancel in the ratio, we can write down simpler
results for the expectation values.  To avoid confusion,  
we shall use the subscript $A$ to denote
the quantities related to the trace in the numerator (i.e.\ depending
on the observable) and  the subscript $0$ for the quantities related
to the denominator, which is independent of the observable.

We also take into account the possibility of having the behaviour of
a part of the original fluctuation spectrum predetermined: If
$\Phi(x)$ describes the known behaviour, we use a new fluctuation
spectrum determined by $f(x)$ which satisfies
$F(x)=f(x)\Phi(x)$ and the observable
$\pref\A=\Phi(\ham)\A$ instead of the original fluctuation spectrum
$F(x)$ and observable $\A$.  Such a separation is sometimes practical
as we shall see in section \ref{sec:tsaex}, or even necessary as we
noted in the beginning of section \ref{sec:PSPA}.

We shall assume in the following
that the Gaussian approximation is valid for the observable $\A$, when
for parameters $L=a(0)$ and $\lambda^2=b^2(0)$,
\begin{equation}\label{e:1stgauss}
\mean{\A}
\approx \frac{ \Tr\!\left[\pref \A G_\lambda
 \!\left(L-g(\ham)\right)\right] }{ 
   \Tr\!\left[\pref G_\lambda\!\left(L-g(\ham)\right)\right] }.
\end{equation}

We shall then apply the PSP-approximation to the Gaussian traces. 
The saddle points $\beta_0$ and $\beta_A$ are determined 
by the equations
\begin{equation}\label{e:gspp}
\beta_0 \lambda^2 + L - \bar{g}_0(\beta_0) = 0\qand
\beta_A \lambda^2 + L - \bar{g}_A(\beta_A) = 0
\end{equation}
and, in general, they are not equivalent.  We estimated in the
preceding section the accuracy of the PSP-approximation for the
Gaussian traces and the bounds derived there can be used for
determining if these approximations are valid.  When this is the 
case, we get from (\ref{e:PSPA}) and (\ref{e:gspp})
\begin{equation}\label{e:PSPAev}
\mean{\A} \approx 
 \left(\frac{\lambda^2+\sigma_0^2(\beta_0)
         }{\lambda^2+\sigma_A^2(\beta_A)} \right)^{\half}  
 \exp\left[\half (\beta_A-\beta_0)^2 \lambda^2\right]
 \frac{ \Tr\!\left[\pref \A 
    \rme^{\beta_A(\bar{g}_0-g(\ham))}\right] }{ 
    \Tr\!\left[\pref\rme^{\beta_0(\bar{g}_0-g(\ham))}\right] },
\end{equation}
where $\bar{g}_0 = \bar{g}_0(\beta_0) = 
{ \Tr\!\left[\pref g(\ham) \rme^{-\beta_0 g(\ham)}\right] 
}/{ \Tr\!\left[\pref \rme^{-\beta_0 g(\ham)}\right] }$.

This result can be simplified for those observables, for which it is
possible to use $\beta_A=\beta_0$.  By the discussion in the previous
section, this is possible at least when
$(\bar{g}_0(\beta_0)-\bar{g}_A(\beta_0))^2\ll\lambda^2+
\sigma^2_A(\beta_0)$, i.e.\ the addition of the observable does not
change the value of $\bar{g}$ significantly.  If also
$\left|\sigma_A^2(\beta_0)-\sigma_0^2(\beta_0)\right| \ll
\lambda^2+\sigma_0^2(\beta_0)$,  the expectation value can be
approximated by the usual canonical formula,
\begin{equation}\label{e:canev}
\mean{\A} \approx  
\frac{ \Tr\!\left[\pref\A \rme^{-\beta_0 g(\ham)}\right] 
   }{ \Tr\!\left[\pref \rme^{-\beta_0 g(\ham)}\right] }.
\end{equation}

\section{Essentially microcanonical systems and
  PSP-entropy}\label{sec:essmc} 

Most applications of statistical methods to
physical systems consider a
large number of particles constrained into a fixed region of
space, either by having them in a container (gases and liquids) with
(potential) walls which prevent the particles from escaping, or by an
attractive interaction between the particles (solids).  In both
cases, the interactions with the environment happen via the boundary
of the container and it is usually plausible to assume 
that in an equilibrium the total energy fluctuations are
negligible.  

This motivates the use of the microcanonical ensemble, 
where energy fluctuations are neglected 
entirely and the density operator is proportional to
$\delta(\ham-E)$.  However, in real systems, there are
interactions with the environment, and although the energy fluctuations
are small, they are not non-existent.  One of the motivations for our
inspection of the Gaussian ensembles was to develop methods for
examination of the effect of these fluctuations. 

We want now to inspect how and when the fluctuations {\em can}\/ be
neglected.
First, we shall assume that the fluctuation spectrum has a compact
support, i.e.\ it is zero outside some finite interval, and we
shall denote its mean and variance by
\begin{equation}\label{e:defeep}
E = \int\! \rmd x F(x) x \qand
  \vep^2 = \int\!\rmd x F(x) (x-E)^2.
\end{equation}
We shall also assume that the reparametrization $g$ is essentially
linear on the support of $F$, i.e.\ that we can use the approximation 
$g(y) \approx g(E) + (y-E) g'(E)$ when computing $\bar{F}$.
This leads to the approximations 
\[
\bar{F}(0) \approx g'(E), \quad a(0) \approx g(E) \qand
  b^2(0) \approx g'(E)^2 \vep^2
\]
which become exact in the limit $\vep\to 0$.  Note also that since the
support of $F$ is compact, we have now $\beta_+=\infty$ and the
parameter range for $\beta$ is thus given by $\beta>\beta_-$.

Thus we can define the parameters of the generalized
Gaussian ensemble by $L=g(E)$ and $\lambda^2=g'(E)^2 \vep^2$, and the
saddle point which yields the best canonical approximation becomes
\[
\beta \lambda^2 + L - \bar{g}(\beta) = 0.
\]
We would like to put $\lambda\to 0$ in this equation, and consider
only solutions to the simpler equation 
\begin{equation}\label{e:betamc}
 g(E) =  \bar{g}(\beta).
\end{equation}
We have seen that this kind of change to the saddle point value is
possible provided the new value satisfies 
$(L+\beta\lambda^2 - \bar{g}(\beta))^2\ll\lambda^2+\sigma^2(\beta)$,
and thus the solution of (\ref{e:betamc}) is a good approximation
if, for example, $\beta^2\lambda^2\ll \beta\sigma(\beta)$.
We assume now that $\vep$ is so small that this approximation can be
made and we shall call such systems essentially microcanonical. 

A straightforward application of the Lebesgue dominated
convergence theorem reveals that
$\lim_{\beta\to\infty}\bar{g}(\beta)=g(E_0)$, where $E_0$ is the
lowest energy for which the expectation value of the observable 
$\A$ is non-vanishing. For example, 
if $\A=\ope{1}$ then $E_0$ is the ground state energy.
Since $\bar{g}$ is continuous and monotonely decreasing,
we can then deduce that a necessary and sufficient condition 
for equation (\ref{e:betamc}) to
have solutions $\beta>\beta_-$ is given by 
$g(E_0) < L < \bar{g}(\beta_-)$ or, if expressed in terms of energy,
\[
E_0 < E < g^{-1}(\bar{g}(\beta_-)).
\]

The microcanonical entropy $\smc$
is defined as the logarithm of the density
of states.  One possible quantitative
definition for the density of states at energy $E$ at the scale
$\vep$, would be given by the number of states in the energy interval
$[E-\half\vep,E+\half\vep]$ divided by the length of the interval, 
$\vep$.  Then the microcanonical entropy $\smc(E,\vep)$ would 
be equal to $\ln\Tr F(\ham)$, where $F$ 
is the fluctuation spectrum proportional to the characteristic 
function of the interval $[E-\half\vep,E+\half\vep]$.
We shall generalize this a bit and inspect the approximation of
$\ln\Tr F(\ham)$ for a general fluctuation spectrum $F$.
The Gaussian (\ref{e:gaussapp}) and saddle point approximations
(\ref{e:PSPA}) yield the estimates
\begin{eqnarray}
\fl \ln \Tr F(\ham) \approx \ln
 \frac{\bar{F}(0)}{\sqrt{2\pi\lambda^2}} +
  \ln \Tr \exp\biggl(-\frac{(L-g(\ham))^2}{2\lambda^2}\biggr) \\
 \lo\approx -\half \ln(2\pi)
 -\half \ln \frac{\lambda^2+\sigma^2}{\bar{F}(0)^2} + 
  \beta\bar{g}(\beta) + \ln \Tr \rme^{-\beta g(\ham)}, \label{e:smcapp}
\end{eqnarray}
where in the second formula we have used $\beta$ defined by equation
(\ref{e:betamc}) for the choice $\A=\ope{1}$. We have also not included
the possible prefactors $\ope{\Phi}$ which where introduced in section
\ref{sec:evs}---these would unnecessarily complicate the
formulae and their inclusion only amounts to the replacements
$\Tr\to\Tr\pref$.

In (\ref{e:smcapp}), the first two terms 
are logarithmic corrections depending on the scale of the
fluctuations and they are usually dominated 
by the remaining terms.
For reasons becoming apparent soon, we shall call
the remaining terms the {\em PSP-entropy,}\/
\begin{equation}\label{e:Spspa}
 \spsp(\beta;\ham) =  \beta 
  \frac{ \Tr\left( g(\ham) \rme^{-\beta g(\ham)} \right)  
    }{\Tr \rme^{-\beta g(\ham)}}
 + \ln \Tr \rme^{-\beta g(\ham)}.
\end{equation}
Comparing this to the definition of 
the canonical Gibbs entropy, $\scan$, shows then that 
$\spsp(\beta;\ham) = \scan(\beta;g(\ham))$, i.e.\ this is the same as
the Gibbs entropy if the Hamiltonian $\ham$ is replaced by 
the reparametrized energy operator $g(\ham)$.

One consequence of this identification is that the maximum entropy
principle holds also for $\spsp$ in the following form
\begin{theorem}[Maximum PSP-entropy] \label{th:maxS}
Suppose $g(\ham)$ is a self-adjoint operator and
$E$ is a real parameter such that equation (\ref{e:betamc}) has a
solution $\beta>\beta_-$.  Then there is a unique positive operator
$\tihop$ which maximizes the Gibbs entropy
functional 
\[
 S[\tihop] = \Tr(-\tihop\ln\tihop)
\]
under the restrictions
\[
\Tr\tihop = 1 \qand g(E) =  \Tr(\tihop g(\ham))
\] 
and this $\tihop$ has the canonical form
\[
\tihop = \frac{ \rme^{-\beta g(\ham)} }{\Tr \rme^{-\beta g(\ham)}}.
\]
\end{theorem}
This theorem follows by an application of the standard result for the
canonical ensemble when the self-adjoint operator $\ham$ is replaced
by $g(\ham)$; for a proof of the standard case, see e.g.\
section 4.2.2 of \cite{Balian1}. 

Define then $\beta(E)$ as the function which maps an energy in the
allowed interval $E_0<E<g^{-1}(\bar{g}(\beta_-))$ to the solution of
equation (\ref{e:betamc}). An application of the implicit function
theorem then shows that $\beta$ is differentiable and that
\[
\beta'(E) = -\frac{1}{g'(E)} \sigma^2(\beta(E)).
\]
Since by equation (\ref{e:Spspa}),
$\rmd\spsp(\beta)/\rmd\beta = -\beta\sigma^2(\beta)$, 
we can conclude that
\begin{equation}\label{e:deds}
\frac{\rmd \spsp(E)}{\rmd E} = \beta g'(E)
\end{equation}
where we abuse
the notation in the usual way and denote $\spsp(E) = \spsp(\beta(E))$.
One immediate conclusion from this equation is that if $\beta_-<0$,
then the inverse $E(S)$ is unique only if the two regions, 
$E<g^{-1}(\bar{g}(0))$ when $\beta>0$ and 
$E>g^{-1}(\bar{g}(0))$ when $\beta<0$, are treated separately.

\subsection{Choosing effective Hamiltonians by the maximum entropy
principle} 

We shall now show that it is possible to use the maximization 
of $\spsp$ for choosing parameters in effective Hamiltonians. 
This method is a simple extension of the standard one---for 
an application of the standard method
see e.g.\ chapter 9.3.\ of the book by Balian \cite{Balian1}.

The easiest way of applying the maximum entropy principle is through
an effective potential $U$, which in the standard case $g=\id$ is
called free energy and denoted by $F$.  Define $U$ by 
\begin{equation}\label{e:defU}
U(T;\ham) = - T \ln \Tr \rme^{-\frac{1}{T}g(\ham)},
\end{equation}
for all $T\in\R$ for which the trace convergences.
A differentiation of this equation shows then that 
\[
\frac{\rmd}{\rmd T} U(T;\ham) \equiv
U'(T;\ham) = -\spsp\biggl(\frac{1}{T};\ham\biggr),
\]
and thus we get the following corollary to theorem \ref{th:maxS},
\begin{corollary}\label{th:effham}
Let $\ham$ be the Hamiltonian, let $g$ be an energy 
reparametrization and let $E$ be a parameter such that
equation (\ref{e:betamc}) has a solution $\beta(E)>\beta_-$ and  
define $T=1/\beta(E)$.  

If $\nham$ is a selfadjoint operator for which 
\begin{equation}\label{e:gecond}
 g(E) = \frac{ \Tr\left( g(\ham) \rme^{-\frac{1}{T} 
	g(\nham)} \right)  
    }{\Tr \rme^{-\frac{1}{T} g(\nham)}}
\end{equation}
then
\begin{equation}\label{e:minup}
 U'(T;\nham) \ge U'(T;\ham)
\end{equation}
and the minimum is attained only for $\nham=\ham$.
\end{corollary}
Note that in (\ref{e:gecond}) the expectation value is of the 
{\em original}\/ Hamiltonian.  

The theorem is usually used in the following manner:
Suppose that the true Hamiltonian $\ham$ is too complicated for the
computation of the trace in $U(T;\ham)$, but there is a
physically plausible effective Hamiltonian
$\hameff$ depending on some parameters $\{x_i\}$ for which the
function $U(T,\{x_i\};\hameff)$ and its $T$-derivative 
$U'(T,\{x_i\};\hameff)$
can be computed. Assume also that it is possible to compute and
invert equation (\ref{e:gecond})
for one of the variables.  
By inserting the solution to $U'(T,\{x_i\};\hameff)$ 
and minimizing the result with respect to the remaining
variables, we get the best possible---in the sense of
entropy---parameters for the 
effective Hamiltonian.

As a curiosity, let us point out that $U(T)$ can also be
obtained from a Legendre transform of the reparametrized energy.
When $\beta_-=0$, the inverse of $\spsp(E)$ is unique and if we denote
it by $E_{\rm psp}(S)$, it is easy to check that $U(T)$ defined by 
\[
U(T) = \inf\!_{S}\{g(E_{\rm psp}(S))-T S\}
\]
is equal to (\ref{e:defU}).
If $\beta_-<0$, the inverse is double valued, and the
Legendre transformation of the first branch will define $U(T)$ for
$T>0$ and transformation of the other branch for $T<0$.

\section{The axioms of Tsallis statistics}\label{sec:tsaax}

In the following sections we shall discuss a non-standard
application of the PSP-approximation: we shall inspect the 
connection between the so called Tsallis statistics and a
PSP-approximation with a certain non-trivial energy
reparametrization.  For this comparison, we shall present 
a few main properties of the Tsallis statistics in this section.

In 1988, C Tsallis proposed \cite{tsa88} a possible generalization of
the principles leading to the Boltzmann statistics.  Motivated by the
usefulness of entropy functionals different from the Gibbs one in the
analysis of fractal phenomena, he proposed a two-parameter
generalization of the Gibbs entropy and expectation values of
observables and then derived a generalization of the Boltzmann 
statistics using the maximum entropy principle.  
In the following years, the
new results where applied to many interesting non-extensive
phenomena, such as stellar polytropes \cite{pla93}, anomalous
diffusion \cite{tsa95,za95} and electron plasma columns
\cite{boh96,hua94}. Especially the results from a 
maximization of a Tsallis entropy functional for the electron 
plasma column were promising, since this
gave results more consistent with experimental observations
than a maximization of a similar Gibbs entropy.

In these applications the principles were set up in the form of
axioms, and the following are cited from \cite{boh96}:
\begin{axiom}[Escort Probabilities]
The system is described by $W$ microscopic state  probabilities
$p_i\ge 0$, which can be used to define a density matrix $\tihop$.
These ``escort probabilities'' satisfy  
$\sum_{i=1}^W p_i = \Tr\tihop = 1$.
\end{axiom}
\begin{axiom}[q-Entropy]
The entropy of the system is defined in terms of two real parameters
$k$ and $q$,
\begin{equation}\label{e:defSq}
S_q[p_i] = k \frac{\sum_{i=1}^W p_i^q -1}{1-q} =  
  k \frac{ \Tr \qtihop-1 }{1-q}.
\end{equation}
\end{axiom}
\begin{axiom}[q-Expectation Value]\label{ax:qexp}
The expectation value of an observable $\A$ which has an expectation
value $a_i$ in the state number $i$ is given by the formula
\[
\mean{\A}_q = \sum_{i=1}^W p_i^q a_i = \Tr (\A\qtihop).
\]
\end{axiom}

With these axioms a maximization of $S_q$ while holding $q$ and  the
$q$-expectation value of the Hamiltonian fixed will yield
\cite{gue96} the canonical Tsallis ensemble
\begin{equation}\label{e:cantsa}
p_i = \frac{1}{ Z_q} \max(0,1-(1-q)\beta_q\vep_i)^{1/(1-q)},
\end{equation}
where $Z_q$ enforces the normalization $\sum p_i =1$ and $\beta_q$
results from using a Lagrange multiplier technique in the
maximization.  It is evident from this formula
that in the limit $q\to 1$ the result reduces to the
standard canonical ensemble.

These axioms are, however, slightly controversial.  As has been noted
before \cite{cza99},  in this axiomatic version we do not have a
proper normalization of the expectation values, since typically
$\mean{1}_q\ne 1$ if $q\ne 1$.  
Here we propose that the easiest solution, the
so called third choice presented in \cite{tsa98}, 
is the correct solution.  
In section \ref{sec:tsaex} we present some possible
conditions under which the canonical Tsallis ensemble would offer a
better description of the system than the canonical Boltzmann
ensemble.  However, for this we find it necessary that 
Axiom \ref{ax:qexp} is replaced by  
\renewcommand{\theaxiomp}{\arabic{axiomp}'}
\setcounter{axiomp}{2}
\begin{axiomp}[q-Expectation Value]\label{ax:qexpp}
The expectation value of an observable $\A$ which has an 
expectation value $a_i$ in the state number $i$ is given 
by the formula
\[
\mean{\A}_q = \frac{\sum_{i=1}^W p_i^q a_i}{\sum_{i=1}^W p_i^q} 
  = \frac{ \Tr\!\left[\A \qtihop\right] }{ \Tr \qtihop }.
\]
\end{axiomp}
We shall discuss the use and meaning of
these axioms in more detail in section \ref{sec:tsaexpl}, but 
first we need to show how the canonical Tsallis ensemble can be
obtained from a PSP-approximation.
 
\section{Tsallis statistics from a PSP-approximation}\label{sec:tsaex}

For the definition of the reparametrization which leads to the Tsallis
statistics we need a new energy parameter, $E_m$.  This
parameter denotes the maximum energy the system can have and we shall
take this into account explicitly by using a prefactor 
$\Phi(E)=\theta(E_m-E)$ in the traces.  Physically, $E_m$ could
represent an energy beyond which the system would
evaporate or, if we are using effective Hamiltonians, beyond which
the effective description is no longer valid.  For now, let $E_m$
simply be a parameter which
bounds the allowed energy range from above.  

Naturally, it is necessary to have $E_m$ greater that the ground state
energy $E_0$ for this kind of restriction to make any sense.  To
simplify the following discussion we now assume that the Hamiltonian
has been normalized so that the ground state energy is equal to 
zero; this can always be accomplished by replacing $\ham$ by $\ham-E_0$.
Then we have $E_m>0$ and we can define 
the reparametrization by $g(E)=-\ln\left(1-E/E_m\right)$.
Applying this to equation (\ref{e:canev}) shows that the
canonical approximation is then given by
\begin{equation}\label{e:tsallispsp}
\mean{\A} \approx  
\frac{ \Tr\!\left[  \A \Phi(\ham)(1-\ham/E_m)^\beta\right] 
   }{ \Tr\!\left[\Phi(\ham) (1-\ham/E_m)^\beta\right] }.
\end{equation}

If we now define $q$ and $\beta_q$ by
\begin{equation}\label{e:defq}
q=\frac{\beta}{\beta+1} \qand 
\beta_q = \frac{1}{(1-q) E_m},
\end{equation}
then the right hand side of (\ref{e:tsallispsp}) becomes  equal to the
canonical  Tsallis expectation values as defined by (\ref{e:cantsa})
and Axiom \ref{ax:qexpp}.  This alone would justify the expression for
the Tsallis expectation values as an approximation to the original
ensemble.  It also gives a connection, via the saddle point equation,
between the parameters of the Tsallis ensemble, $q$ and $\beta_q$, and
the characteristics of the original ensemble, $E_m$, $L$ and
$\lambda^2$.  It does not, however, tell us when this approximation
would be better than the usual canonical one.  We shall come back to
this in the following section.

In order to get some feeling how the saddle point approximation 
behaves in this case,
we shall now assume that the density of states increases
polynomially and in the computation of the traces we shall 
apply the approximation 
\begin{equation}\label{e:poldens}
\Tr f(\ham) \simeq c
 \int_{0}^\infty\!\! \rmd x\, x^{n-1} f(\omega x),
\end{equation}
where $\omega$ defines the relevant energy scale for the increase of
the density of states and $c$ is a constant independent of $f$.  
This approximation is not quite as arbitrary as it looks, since it
can be derived e.g.\ from the high-energy behaviour of
a system of $n$ harmonic oscillators.
For energy observables $\A=\ham^k$ equation (\ref{e:poldens}) yields
\begin{eqnarray}\label{e:tsagamma}
\fl
\Tr\!\left[\pref\ham^k\rme^{-\beta g(\ham)}\right] \simeq 
 c (E_m/\omega)^n E_m^k
  \int_0^1\! \rmd y\, y^{n+k-1} (1-y)^\beta 
  \propto E_m^{k} 
  \frac{\Gamma(n+k)\Gamma(\beta+1)}{\Gamma(n+k+\beta+1)}.
\end{eqnarray}
Clearly, for all $k\ge 0$ we now have $\beta_-=-1$
and thus we need not worry about the applicability of the
Gaussian approximation for large scale observables.

We could now examine the accuracy of the canonical approximation to
the Gaussian traces by using the bounds derived in section
\ref{sec:canacc}. 
Going through these computations, however, is quite tedious and
probably not very interesting, so we shall give a few
numerical examples instead.

By the approximation (\ref{e:tsagamma}), 
the canonical expectation values are now
\begin{equation}\label{e:tsamean}
\mean{\ham^k}_\beta \simeq E_m^k \prod_{j=0}^{k-1}
  \frac{1}{1+\frac{\beta+1}{n+j}}.
\end{equation}
where the parameter $\beta$ is determined from the saddle point
equation involving the digamma function 
$\psi(x)=\frac{\rmd}{\rmd x}\ln \Gamma(x)$,
\begin{equation}\label{e:tsasp}
L + \beta \lambda^2 = \psi(n+\beta+1)- \psi(\beta+1).
\end{equation}
For instance, the expectation value and the variance of energy are then
\begin{equation}\label{e:tsavar}
p\equiv \frac{\mean{\ham}}{E_m} =
\frac{1}{1+\frac{\beta+1}{n}}\qand \mean{\ham^2} - \mean{\ham}^2 
= \frac{1-p}{n+p} \mean{\ham}^2
\end{equation}
and the percentage of fluctuations of the total energy in the
canonical ensemble is, therefore,
approximately $\sqrt{(1-p)/n}$.  If we had used the Boltzmann
ensemble here, the total energy fluctuations would have been given by
$1/\sqrt{n}$, and thus they would have been larger than in the Tsallis
case by the factor of $\sqrt{1-p}$.

We have given a few examples of what kind of
errors the coarse graining and the Gaussian, PSPA and canonical
approximations induce to an initially microcanonical fluctuation
spectrum in table \ref{t:tsares}.  
Of course, the Gaussian approximation is closest to the
microcanonical case when $\lambda\ll 1$ and the closer $E/E_m$ is to
1, the larger values of $\lambda$ can be used before
the energy expectation value changes significantly.  
The canonical Tsallis
approximation has the expected behaviour, but the accuracy of the
PSPA-result is surprisingly good, especially at reproducing the energy
fluctuations of the Gaussian ensemble.  
We also want to point out that $\beta$ can have negative values here
even though this makes the density operator apparently singular.

\begin{table}
\caption{The effect of coarse graining and the computation of 
expectation values in the different approximations: the 
first row gives the expectation value of energy, 
$\mean{\ham}/E_m$, and the second row the percentage of 
energy fluctuations, $\text{Var}(\ham)^{1/2}/\mean{\ham}$.  
The three expectation values refer to equations 
(\ref{e:1stgauss}), (\ref{e:PSPAev}) and (\ref{e:canev}), 
respectively.\label{t:tsares} }
\begin{indented}
\lineup
\item[]\begin{tabular}{@{}*{10}{l}}
\br 
\centre{3}{Parameters} & & & & & \centre{3}{Expectation values} \\
\crule{3} & & & & & \crule{3} \\  
\0\0\( n \)& \( E/E_{m} \)&  \( \lambda  \)& 
 \0\0\( \beta \) & \0\0\( \beta_{A} \)& 
 \( \lambda/\sigma_0 \)& \m\( q \)& \( G_{\lambda } \)
 & PSPA & Tsallis\\
\mr  
\0\05&0.1&0.05&\026.3&\029.9&0.65&\m0.96&0.1503&0.1507&0.1548\\
&&&&\033.3&&&0.2345&0.2329&0.4049\\
\0\05&0.5&0.5&\0\01.81&\0\02.05&0.93&\m0.64&0.6398&0.6422&0.6403\\
&&&&\0\02.26&&&0.1912&0.1885&0.2525\\
\0\05&0.99&0.1&\0\0\-0.66&\0\0\-0.63&0.03&$-1.90$&0.9899&0.9843&0.9355\\
&&&&\0\0\-0.61&&&0.0010&0.0336&0.1043\\
\0\05&0.99&5.0&\0\0\-0.09&\0\0\-0.08&3.84&$-0.10$&0.8450&0.8450&0.8456\\
&&&&\0\0\-0.08&&&0.1603&0.1603&0.1625\\
100&0.1&0.05&155.7&156.5&1.00&\m0.99&0.3891&0.3891&0.3895\\
&&&&157.3&&&0.0554&0.0554&0.0780\\
100&0.5&0.05&\071.4&\072.0&0.56&\m0.99&0.5806&0.5806&0.5801\\
&&&&\072.5&&&0.0316&0.0316&0.0646\\
100&0.99&0.1&\0\00.47&\0\00.48&0.10&\m0.32&0.9900&0.9901&0.9855\\
&&&&\0\00.49 &&& 0.0010 & 0.0008 & 0.0120\\ 
\br
\end{tabular}
\end{indented}
\end{table}

\section{Towards Tsallis thermodynamics}\label{sec:tsaexpl}

Consider the following experiment: take a macroscopic piece of
material in thermal equilibrium and separate a small sample from
it. Let the sample be contained in some finite volume by a finite
potential and wait until it again reaches an equilibrium.   Now the
sample will have lost energy and possibly particles by evaporation
from the surface, especially there will be no particles  with enough
kinetic energy to escape the potential barrier in the sample. If the
original temperature was high enough, the average energy density would
also decrease and one would expect the system to end up in a state
with average energy per particle near the escape potential and with a
sharp drop for higher energies.  If the number of particles in the
final sample is small enough, the broad high-energy tail  indicated by
the canonical ensemble will give an observable change to the predicted
behaviour of the system.  The example in section
\ref{sec:tsaex} shows that under these
conditions the Tsallis statistics with a suitable choice of $E_m$ can
offer a better alternative.

There are also instances when the canonical ensemble is ill-defined,
but we can still resort to the Tsallis ensemble with its finite energy
cut-off.  If the density of states increases faster than exponentially,
the canonical ensemble cannot be used, whereas the Tsallis ensemble
will always produce finite expectation values.  Another, more
surprising, example is given by systems with a ``Coulomb-like''
spectrum: these systems have a point spectrum only below certain value
$E_c$ while the spectrum above $E_c$ is continuous or, what is
relevant to our case, does not contain any eigenvalues.  For these
systems any trace-class operator of the form $F(\ham)$ must have a
cut-off at the energy $E_c$---thus canonical ensemble will not work,
but Tsallis ensemble with $E_m< E_c$ is well-defined.  

We now wish to examine what happens if
the fluctuation spectrum $F$ is essentially
microcanonical, i.e.\ when the total energy fluctuations are small
enough that the reparametrization  
$g$ is essentially linear on the support of $F$ and 
we can use the saddle point values from the approximate equation
(\ref{e:betamc}).  Let $E$ denote the expectation value 
of the distribution $F$, when necessarily $E<E_m$, and let
$\beta(E)$ be the solution to the approximate saddle point equation,
\begin{equation}\label{e:deftsbeta}
\ln(1-E/E_m) = \frac{\Tr\biggl[\ln(1-\ham/E_m) \theta(E_m-\ham)
  (1-\ham/E_m)^\beta\biggr]}{
  \Tr\biggl[\theta(E_m-\ham) (1-\ham/E_m)^\beta\biggr]}.
\end{equation}

The Tsallis parameters $q$ and $\beta_q$
are then obtained from $\beta$ and $E_m$ as in
the previous section, by equation (\ref{e:defq}).
We shall also adopt the notation
\begin{equation}\label{e:defescm}
\escm = \left(1-(1-q)\beta_q \ham\right)^{\frac{1}{1-q}} 
  \theta(1-(1-q) \beta_q \ham),
\end{equation}
when the escort matrix defined in section \ref{sec:tsaax} is 
given by $\tihop=\escm/\Tr\escm$.

Let us then assume that the canonical approximation is accurate for
the energy observable:  $E\simeq\Tr(\ham\qtihop)/\Tr\qtihop$.  
In the examples in table \ref{t:tsares} this is the case 
whenever the energy $E$ lies near the cutoff $E_m$.
Using the definition (\ref{e:defescm}), we can then deduce
that
\begin{equation}\label{e:eemratio}
1-\frac{E}{E_m} = \frac{\Tr \escm}{\Tr\escm^q}.
\end{equation}

We showed in section \ref{sec:essmc} that the PSP-entropy is 
an approximation to the logarithm of the density of states,
$\ln W$.  
Applying the definition of $\beta$ and of 
the Tsallis parameters then yields the formula
\begin{equation}\label{e:lnW}
\ln W \approx
\spsp = -\frac{q}{1-q} \ln(1-E/E_m) + \ln \Tr \escm^q
 = \frac{1}{1-q} \ln 
  \biggl[ \frac{\Tr \escm^q}{(\Tr\escm)^q}\biggr],
\end{equation}
where in the last equality we have used (\ref{e:eemratio}).
Thus the $q$-entropy satisfies for $\tihop=\escm/\Tr\escm$,
\begin{equation}\label{e:sqw}
S_q[\tihop] = \frac{1}{1-q} 
  \left( \frac{\Tr\escm^q}{(\Tr\escm)^q} -1 \right) 
\approx \frac{ 1}{1-q} \left( W^{1-q} -1 \right).
\end{equation}
This relation between the density of states and the Tsallis-entropy
was one of the motivations used in the original article
\cite{tsa88} for the definition of $S_q$.

The other results proven in section \ref{sec:essmc} apply here as
well, especially we can conclude that
the energy dependence on changes of $\beta$ 
while holding $E_m$ fixed satisfies
\begin{equation}\label{e:depspa}
\frac{\partial}{\partial\beta} \spsp = \frac{\beta/E_m}{1-E/E_m}
  \frac{\partial}{\partial\beta} E.
\end{equation}
Interestingly, the Tsallis parameters yield a similar equation if we
fix $q$ and let $\beta_q$ vary.  For this, define $\beta_q(E,q)$ as the
solution to the equation 
\[
 E = \frac{\Tr(\ham \escm^q)}{\Tr\escm^q}.
\]
Then a straightforward but lengthy computation shows that 
\[
\frac{\partial}{\partial\beta_q} \frac{\ln(1+(1-q)S_q)}{1-q}
 = \frac{\beta_q}{1-(1-q) \beta_q E}
\frac{\partial}{\partial\beta_q} E
\]
which by (\ref{e:lnW}) and (\ref{e:sqw}) is essentially the same
equation as in (\ref{e:depspa}).

These results clarify the difference between the use of 
the Tsallis parameters and the PSP-parameters.  In the first case,
the increase of the fluctuation spectrum, defined by $q$, is a known
quantity, while the boundary conditions determine the position of the
energy cut-off, defined by $\beta_q$.  In the PSP-approximation, 
it is more natural to vary the energy of the system
while holding the energy cut-off fixed.  Otherwise, these two methods
are very similar, as both have a simple formula for the change of
energy and an maximum entropy principle which can be applied for a
determination of effective Hamiltonians, as explained in section
\ref{sec:essmc}.

In real experiments performed on a sample of
matter in a container, one usually examines the properties of the
sample matter by altering some easily controllable
physical conditions of the environment.  
For example, if the temperature of the container is the varying quantity, 
then a natural parameter for the experiment
would be the energy density of the sample matter at the boundary.
Unfortunately, unless the state of the system is homogeneous, 
the energy density on the boundary is not determined trivially
from the total energy given by the Hamiltonian.
It is, of course, possible that the
boundary conditions can be included naturally to the dynamics by using
effective Hamiltonians and this would then
determine the relations between the
parameters of the experiment and the parameters of the
PSP-approximation. 

There has already been a lot of discussion of the physical meaning of 
the Tsallis parameter $q$ in the literature.   For
example, for systems exhibiting a multifractal behaviour the
value of $q$ could be quite straightforwardly determined 
as was shown in \cite{lyra98}.  The value of $q$ can be also related
to the degree of non-extensivity of energy and the speed of growth of 
certain classical long-range potentials \cite{jund95}.

\section{Speculations about the maximum entropy 
  principle}\label{sec:maxS}

As mentioned earlier, the Tsallis ensemble is usually derived by
postulating the Tsallis entropy functional (\ref{e:defSq}) and then
finding its maximum under the condition that the energy expectation
value is fixed.  This approach is motivated by the standard derivation
of the canonical ensemble by a similar maximization of the Gibbs entropy
functional.

The motivation behind these derivations lies in the intuitive
association of the maximum likelihood ensemble with the least
restricted, maximum disorder, ensemble.  Since for systems with
infinitely many degrees of freedom the natural choice of associating
the maximum likelihood with equiprobability is not possible, it is
then necessary to postulate an entropy functional which quantitatively
measures the disorder of any density operator.  The Gibbs entropy
functional $S[\tihop]=\Tr(-\tihop\ln\tihop)$ can be derived by
requiring the entropy to be additive for independent systems.  The
additivity of the thermodynamical entropy, on the other hand, is
closely related to the extensivity of thermodynamical 
systems---see, for example, the axiomatic derivation 
of Lieb and Yngvason \cite{lieb99} which clearly
highlights the role of extensivity.
Since the Tsallis entropy is only sub- or superadditive \cite{boh96}, 
by this reasoning it would be useful only for systems which are
non-extensive.  

In this work we have adopted a different implementation of the maximum
disorder principle:  by coarse graining of energy we try to find a
simple {\em parametrization}\/ of the underlying ensemble which would
have all the same relevant predetermined properties as the original
ensemble.  The coarse graining will naturally increase the number of
energy states which participate in the ensemble and thus we are, in a
way, always increasing the microcanonical entropy of the ensemble.

We now claim that in spite of the similarities, the present approach
is conceptually easier and more readily applicable to a wider variety
of situations.  First, the possible
fluctuations of the parameters can also be
properly taken into account in this formulation and the effect of
these fluctuations can be analysed. Secondly, although the energy
reparametrization $g$, which is responsible for the appearance of 
non-standard ensembles, plays the same role as the
different entropy functionals in the maximum entropy derivations, its
interpretation is more tangible and, consequently, choosing between
different non-standard ensembles should be easier.

We would also like to emphasize the role of entropy
differently here.  Instead of a fundamental,
philosophical role the entropy has been endowed, we would like to
emphasize its {\em practical}\/ importance both as a function encoding
the dependence of energy on the temperature-parameter and as an
effective potential which enables one to choose parameters in effective
descriptions of the system.  

As we have shown, our definition of
PSP-entropy, (\ref{e:Spspa}), leads to an 
extension of both of these properties.  The relation between 
the changes of total energy and PSP-entropy are given in
(\ref{e:deds}), from which the standard case follows by setting
$g=\id$. The maximization of the statistical
PSP-entropy was proven in theorem \ref{th:maxS} and its corollary
provides a way of choosing between effective
Hamiltonians.  The Tsallis axioms also offer another, 
slightly different,
way of applying the maximization of entropy which 
leads to one of the canonical PSP-ensembles.

\section{Conclusions}

We have presented a generalized coarse graining procedure which can be
used for analyzing a wide variety of quantum systems.  The main
purpose of the coarse graining was to develop simple parametrizations
which could be used for analyzing the behaviour of the system when its
energy is varied.  We restricted to the energy only, but the method can
be easily extended to the case when there are many relevant conserved
quantities.

One important application was to systems for which the Boltzmann
ensemble either does not converge, or produces too large total energy
fluctuations.  We have shown that in this case the Tsallis statistics
offers a generalization which is more likely to work, but which still
retains some of the useful properties of classical thermodynamics: 
maximization of an  entropy functional and a simple formula
for the differential change of energy.  The cost in this case is the
inclusion of an extra parameter which needs to be chosen in a
suitable way dependent on the behaviour of the system---the traces
involved in the computations are also likely to be prohibitively
difficult to compute.

In the final section we have raised some objections to the use of the
maximum entropy principle as a fundamental justification of the
canonical Boltzmann ensemble.  Especially for purposes of physics
teaching, it has the disadvantage that it either leads to the idea
that canonical ensemble is the only sensible statistical description
of an arbitrary physical system or that thermodynamics can only be
applied in the thermodynamical limit in which all the standard
ensembles agree.  We consider both of these ideas to be 
misguidingly restrictive.

The general guidelines for choosing an ensemble could be the
following: the number of parameters should be as few as possible
and have as direct as possible relation to the 
physical characteristics of the system. In addition, the
traces needed for the evaluation of the ensemble should be computable
for the given Hamiltonian and, naturally, the ensemble must reproduce
the original ensemble within the required accuracy.  
Of the coarse grained ensembles we considered,  it should
always be possible to find a Gaussian ensemble which is 
accurate enough.
The Gaussian expectation values, however, are usually difficult to
evaluate, while the corresponding canonical ensembles are often easier
in this respect.   Of the canonical ensembles, the Boltzmann ensemble
is the simplest but, when it cannot be used, the canonical
Tsallis ensemble is a good second trial.

There is, however, one more possibility we have not stressed yet.
From table \ref{t:tsares} we can see that the PSP-approximation of
expectation values, equation (\ref{e:PSPAev}), can give significantly
better approximation to Gaussian expectation values than the
canonical ensemble.  On the other hand, for the simple case $g(x)=x$,
everything needed in the PSP-approximation 
formula is given by expectation values
in the usual canonical ensemble.  Using the PSP-approximation 
would thus enable the
approximation of Gaussian expectation values without having to
compute anything but canonical expectation values---this might be
especially useful for small samples of macroscopic matter for which
the canonical expectation values are computable, but non-accurate.

\ack

The author warmly thanks C.~Tsallis for his help 
and wishes to express gratitude for Tsallis and
I.~Procaccia for a stimulating discussion.  I am also indebted to
A.~Kupiainen for valuable discussions.

\appendix

\section{Proof of the Gaussian approximation theorem}\label{sec:gausspf} 

We shall begin the proof of theorem \ref{th:gaussconv} from 
equation (\ref{e:trFL}) which was shown in section \ref{sec:gaussapp}
to follow from the assumptions \ref{it:gobs} and \ref{it:gexpdecay}. 
In other words, we inspect the integral representation 
\begin{equation}\label{e:apptrFL}
\Tr\bigl(\A F_\Lambda(\ham)\bigr)  = 
 \int_{\beta-\ci\infty}^{\beta+\ci \infty}  
  \frac{\rmd w}{ 2\pi\ci} \rme^{\half\Lambda^2 w^2}  
  \bar{F}(w) \Tr\bigl(\A \rme^{-w g(\ham)}\bigr),
\end{equation}
which is valid for all $\beta_-<\beta<\beta_+$ and $\Lambda\ge 0$ 
with an integrand analytic in the strip $\beta_-<\re w<\beta_+$.
The proof will depend on the properties of the 
real saddle point approximation in the large
$\Lambda$ limit so we shall now first examine the large $\Lambda$
dependence of the real saddle point value.

By differentiation of the logarithm of the integrand we arrive at
the saddle point equation
\begin{equation}\label{e:lsp}
\Lambda^2 w + a(w) - \bar{g}(w) = 0,
\end{equation}
where the functions $a$ and $g$ are defined in (\ref{e:defpars}).
A differentiation of the left hand side yields then
$\Lambda^2+b^2(w)+\sigma^2(w)$ which is clearly strictly
positive for real $w$.
Thus the restriction of the left hand side to the real axis is
strictly increasing and we have a real saddle point $\beta$
on the interval $\beta_-<\beta<\beta_+$ if and only if
\begin{equation}\label{e:isbeta}
\Lambda^2 \beta_- + a(\beta_-) < \bar{g}(\beta_-) \qand 
\Lambda^2 \beta_+ + a(\beta_+) > \bar{g}(\beta_+).
\end{equation}
Since $\beta_+>0$, the second inequality is satisfied as soon as
$\Lambda$ becomes large enough.  If $\beta_-<0$, then the same holds
for the first inequality as well, but if $\beta_-=0$, then we need to
have $a(0) < \bar{g}(0)$ which is equivalent to the condition
\ref{Git:issp} of section \ref{sec:gaussapp}.

Let us now assume that either of these conditions hold and let
$\Lambda_c\ge0$ be the infimum of values for which
(\ref{e:isbeta}) is valid.  Then for all $\Lambda> \Lambda_c$ there is
a unique real saddle point which we denote by $\beta(\Lambda)$.  As
the left hand side of (\ref{e:lsp}) is strictly increasing, we can also
deduce that $\beta(\Lambda)$ never changes sign:
\begin{enumerate}
\item if $a(0)<\bar{g}(0)$, then $\beta(\Lambda)>0$,
\item if $a(0)=\bar{g}(0)$, then $\beta(\Lambda)=0$,
\item if $a(0)>\bar{g}(0)$, then $\beta(\Lambda)<0$,
\end{enumerate}
for all $\Lambda>\Lambda_c$.  

An application of the implicit function theorem to the defining equation
(\ref{e:lsp}) proves that $\beta(\Lambda)$ is a smooth function 
from $(\Lambda_c,\infty)$ to $(\beta_-,\beta_+)$.  A differentiation
of the defining equation then shows that
\[
\beta'(\Lambda) = - \frac{2\beta\Lambda}{
  \Lambda^2+b^2(\beta)+\sigma^2(\beta)},
\]
from which we can deduce that $\beta$ is
either monotonely decreasing positive ($a(0)<\bar{g}(0)$),
zero ($a(0)=\bar{g}(0)$) or monotonely increasing negative
($a(0)>\bar{g}(0)$) function of $\Lambda$.  In all three 
cases, the limit 
$\beta_\infty = \lim_{\Lambda\to\infty} \beta(\Lambda)$ exists and
belongs to the interval $(\beta_-,\beta_+)$ if $\beta_-<0$, and to the
interval $[0,\beta_+)$ if $\beta_-=0$.

We will next show that $\beta_\infty=0$.  Assume, to get a
contradiction, that $\beta_\infty\ne0$.  By the above results, then
$\beta_\infty$ always belongs to the interval $(\beta_-,\beta_+)$ and
thus both $\lim_{\Lambda\to\infty} \bar{g}(\beta(\Lambda))$ and
$\lim_{\Lambda\to\infty} a(\beta(\Lambda))$ exist and are finite.
But taking the limit $\Lambda\to\infty$ in the defining equation
(\ref{e:lsp}) then shows that we must have $\beta(\Lambda) \to 0$,
which is a contradiction.

Thus we have shown that the saddle point solution goes to zero
smoothly and monotonely as the scale $\Lambda$ is increased.  Since we
are only interested in the large scale behaviour, we shall next expand 
$\ln\bar{F}$ in a  neighbourhood of origin; this is possible since 
$\bar{F}$ is analytic and $\bar{F}(0)\ne 0$.  The expansion yields
\begin{equation}\label{e:Ftaylor}
\bar{F}(w) = \bar{F}(0) \rme^{L w + \case12 \lambda^2 w^2}
	(1+\Delta\bar{F}(w)), 
\end{equation}
where $L=a(0)$, $b^2=\lambda^2(0)$ and
the correction term $\Delta\bar{F}$ satisfies 
\begin{equation}\label{e:DFlim}
\lim_{w\to 0} \frac{\Delta\bar{F}(w)}{w^k} < \infty,
\end{equation}
for $k=3$.  

Let us then consider the expansion (\ref{e:Ftaylor}) when 
$L$ and $\lambda^2$ are not necessary equal to the default values
$a(0)$ and $b^2(0)$.  It is easy to deduce
that if $\lambda^2 \ne b^2(0)$, (\ref{e:DFlim}) is true only for 
$k\le 2$ and if we have 
$L\ne a(0)$, then (\ref{e:DFlim}) holds only for $k\le 1$.
However, as is evident from the definition (\ref{e:Fbardef}),
$|\bar{F}(w)|\le\bar{F}(\re w)$ always, 
and we get from the definition (\ref{e:Ftaylor}) a bound
\begin{equation}\label{e:fb}
|\Delta\bar{F}(\beta+\ci\alpha)|\, \rme^{-\half\alpha^2\lambda^2}
\le 1 + \frac{ \bar{F}(\beta) }{
   \bar{F}(0) \rme^{\beta L +\half \beta^2 \lambda^2 } },
\end{equation}
which is valid for all $\beta<\beta_+$ and all real $\alpha$,
$L$ and $\lambda^2$.

For the proof of the theorem, we need to consider the difference
\begin{eqnarray}
\fl
 \Delta_G \equiv \Tr\bigl(\A F_\Lambda(\ham)\bigr) -
  \bar{F}(0) \Tr\left[\A
  G_{\sqrt{\lambda^2+\Lambda^2}}(L-g(\ham))\right] \nonumber \\
 \lo= \int_{\beta-\ci\infty}^{\beta+\ci \infty}  
  \frac{\rmd w}{ 2\pi\ci}
  \Tr\bigl(\A \rme^{-w g(\ham)}\bigr)
  \bar{F}(0) \rme^{L w+\half(\lambda^2+\Lambda^2) w^2}
  \Delta\bar{F}(w), \label{e:deltag}
\end{eqnarray}
where we have used the $g$-transform of the Gaussian trace, the
integral representation (\ref{e:apptrFL}),
and defined the function $\Delta\bar{F}(w)$ 
by equation (\ref{e:Ftaylor}).
For each $\Lambda\ge 0$, let us parametrize the
integration variable in (\ref{e:deltag}) as
$w=\beta+\ci\alpha$.  To prove the convergence properties stated in
the theorem, we need to choose $\beta(\Lambda)$ so that it decays 
like a negative power of $\Lambda$: we assume that $c\ge 0$ and 
that $\beta$ has been chosen so that
\begin{equation}\label{e:betadec}
 \beta^{1+c} \Lambda \mbox{ stays bounded in the limit }
   \Lambda\to\infty.
\end{equation}
In particular, this means that $\beta\to 0$ when $\Lambda\to\infty$.

The absolute value of the correction term then has an upper bound 
\begin{equation}\label{e:corrbound}
\fl |\Delta_G| \le
 \bar{F}(0) \rme^{\beta L +\half \beta^2 (\lambda^2+\Lambda^2) }  
 \Tr\left[\A\rme^{-\beta g(\ham)}\right]
 \int_{-\infty}^\infty \frac{\rmd\alpha}{2\pi} 
 \rme^{-\half (\lambda^2+\Lambda^2) \alpha^2} 
 |\Delta\bar{F}(\beta+\ci\alpha)|
\end{equation}
and we shall inspect the behaviour of the remaining integral next.

Let $k$ be such that (\ref{e:DFlim}) is true; as was mentioned
earlier, we can always choose at least any $k\le 1$.  Then 
there are constants $M>0$ and $m>0$ such that for all $|w|\le 2 m$,
\[
 |\Delta\bar{F}(w)| \le M |w|^k.
\]
Assume also that $\Lambda$ is so large that
$|\beta|<m$.  Then for all $|\alpha|<m$, 
\[
 \Lambda^{\frac{k}{1+c}}
 |\Delta\bar{F}(\beta+\ci\alpha)| \le M 
  |\beta\Lambda^{\frac{1}{1+c}} +\ci(\alpha\Lambda)
       \Lambda^{-\frac{c}{1+c}}|^k 
  \le M (|\beta|\Lambda^{\frac{1}{1+c}} + |\alpha|\Lambda)^k,
\]
which implies that 
\[
 \Lambda^{\frac{k}{1+c}}  \sqrt{2\pi(\lambda^2+\Lambda^2)} 
 \int_{-m}^{m} \frac{\rmd\alpha}{2\pi} 
 \rme^{-\half (\lambda^2+\Lambda^2) \alpha^2} 
 |\Delta\bar{F}(\beta+\ci\alpha)|
\]
stays bounded in the limit $\Lambda\to\infty$.
In fact, the same is then true also when the integration
limits are replaced by $\pm\infty$, 
since the remaining integral over the values $|\alpha|\ge m$
does not contribute at all to the limit $\Lambda\to\infty$;
to see this, apply dominated convergence theorem to the remainder. 

By equation (\ref{e:corrbound}) this proves that 
\begin{eqnarray}\label{e:dgineq}
\fl \frac{|\Delta_G|}{\bar{F}(0) \Tr\bigl[\A
  G_{\sqrt{\lambda^2+\Lambda^2}}(L-g(\ham))\bigr]} \le 
 \frac{\rme^{\beta L +\half \beta^2 (\lambda^2+\Lambda^2) }  
 \Tr\bigl[\A\rme^{-\beta g(\ham)}\bigr]}{
  \Tr\bigl\{\A\exp\bigl[{-\frac{1}{2(\lambda^2+\Lambda^2)}
  (L- g(\ham))^2}\bigr]\bigr\}}
 \order{\Lambda^{-\frac{k}{1+c}}},
\end{eqnarray}
where $k$ and $c$ are any values allowed by (\ref{e:DFlim}) and
(\ref{e:betadec}), respectively.   In the final part of the proof 
we derive conditions when the remaining multiplicative term stays
bounded for large $\Lambda$.

Define first an auxiliary variable $x$ by the equation
$x = L + \beta (\lambda^2 + \Lambda^2)$.
Applying this in (\ref{e:dgineq}), reveals that the multiplicative 
term is the inverse of 
\begin{equation}\label{e:lookforb}
  \frac{ \Tr\bigl[\A\rme^{-\beta g(\ham)}
  \rme^{-\frac{1}{2(\lambda^2+\Lambda^2)}(x- g(\ham))^2}\bigr]}{
   \Tr\bigl[\A\rme^{-\beta g(\ham)}\bigr]}.
\end{equation}
Since (\ref{e:lookforb}) is always less than one, it cannot improve 
the convergence of the right hand side in (\ref{e:dgineq}). 
Thus it is enough to show that it does not spoil the convergence for
the optimal choice $L=a(0)$.
 
Since $\A$ is positive, Jensen's inequality can be applied
in (\ref{e:lookforb}) and we obtain that the logarithm of
(\ref{e:lookforb}) is always greater than or equal to 
\begin{equation}\label{e:Jensenb}
\fl
-\frac{1}{2(\lambda^2+\Lambda^2)} 
  \frac{\Tr\bigl[\A\rme^{-\beta g(\ham)}(x- g(\ham))^2\bigr]}{
   \Tr\bigl[\A\rme^{-\beta g(\ham)}\bigr]} =
 -\frac{(x-\bar{g}(\beta))^2}{2(\lambda^2+\Lambda^2)} 
 -\frac{\sigma^2(\beta)}{2(\lambda^2+\Lambda^2)}.
\end{equation}
Both of the terms in (\ref{e:Jensenb}) are negative and thus they 
have to be bounded separately for the boundedness of the 
multiplicative term in (\ref{e:dgineq}).  

The first term is small only if $\beta$ satisfies 
\begin{equation}\label{e:closebeta}
|L + \beta (\lambda^2 + \Lambda^2)-\bar{g}(\beta)| \lesssim 
 \sqrt{\lambda^2 + \Lambda^2}.
\end{equation}
If we multiply this by $\beta^{1+2 c}$ and take the limit
$\Lambda\to\infty$, it becomes evident that the conditions 
(\ref{e:closebeta}) and (\ref{e:betadec}) can be compatible only if
$\beta^{1+2 c} \bar{g}(\beta)$ stays bounded in the limit 
$\beta\to 0$, i.e.\ only if \ref{Git:gbarcond} holds with the same
value of $c$.

The converse is also true in the sense that if $c\ge 0$ is such that
\ref{Git:gbarcond} holds, then we can find $\beta(\Lambda)$ which
satisfies both (\ref{e:closebeta}) and (\ref{e:betadec}).  
Such a $\beta$ is defined by the equation 
\begin{equation}\label{e:appgsp}
L + \beta (\lambda^2 + \Lambda^2)-\bar{g}(\beta) = 0,
\end{equation}
as we shall now show.  A brief reflection shows that for a Gaussian
fluctuation spectrum 
$F(\ham)=G_{\sqrt{\lambda^2+\Lambda^2}}(L-g(\ham))$, 
we would have $a(w) = L + w (\lambda^2+\Lambda^2)$ 
and thus (\ref{e:appgsp}) is 
the saddle point equation for a Gaussian fluctuation
spectrum.  By the results proven in the beginning of this section, we
then know that if $\beta_-<0$ or $L<\bar{g}(0)$, (\ref{e:appgsp}) 
defines for all large enough $\Lambda$
a smooth function $\beta(\Lambda)$ which monotonely approaches 
zero as $\Lambda$ increases.  Such $\beta(\Lambda)$ trivially
satisfies (\ref{e:closebeta}) and by multiplying (\ref{e:appgsp}) by 
$\beta^{1+2 c}$ we can see that \ref{Git:gbarcond} implies then that
also (\ref{e:betadec}) is true for the same value of $c$.  

Since it is
part of the assumptions so far that either $\beta_-<0$ or
$a(0)<\bar{g}(0)$, we have thus shown that if \ref{Git:gbarcond}
holds, then---at least 
for the optimal choice $L=a(0)$---it is possible to define
$\beta(\Lambda)$ by (\ref{e:appgsp}) and this function satisfies
the decay condition (\ref{e:betadec}).  Also note that
if $\beta_-<0$, then
$\bar{g}(0)<\infty$ and \ref{Git:gbarcond} is trivially satisfied for
the best allowed choice $c=0$.

For the boundedness of the second term in (\ref{e:Jensenb}), 
we shall need either
$\beta_-<0$ or assume the last of the
three conditions, \ref{Git:sigmacond}.  If $\beta_-<0$, then
$\sigma^2(0)<\infty$ and the second term vanishes in the limit
$\Lambda\to\infty$.  If $\beta_-=0$ and $L<\bar{g}(0)$,
we can define $\beta$ by equation (\ref{e:appgsp}) when 
\[
 \frac{\sigma^2(\beta)}{\lambda^2+\Lambda^2} =
   \frac{\beta\sigma^2(\beta)}{\bar{g}(\beta)-L},
\]
and clearly it is then sufficient that
$\beta\sigma^2(\beta)/\bar{g}(\beta)$ stays bounded in the limit
$\beta\to 0^+$. Note that this proves the sufficiency of 
condition \ref{Git:sigmacond} for the relevant case $L=a(0)$.

Combining the results we get then from 
equations (\ref{e:deltag}) and (\ref{e:dgineq}) that if either
$\beta_-<0$ or \ref{Git:first}--\ref{Git:last} hold, then
for $L=a(0)$ and $\lambda^2=b^2(0)$ we have
\[
 \Tr\bigl(\A F_\Lambda(\ham)\bigr) = 
  \bar{F}(0) \Tr\left[\A 
	G_{\sqrt{\lambda^2+\Lambda^2}}(L-g(\ham))\right] 
  \left( 1 + \order{\Lambda^{-\frac{k}{1+c}}} \right)
\]
with $k=3$.  We have also seen that changing $\lambda$ would allow
using only $k=2$ and changing $L$ only $k=1$ or nothing at all---if 
$\beta_-=0$ and $L> \bar{g}(0)$, there are no solutions to equation
(\ref{e:appgsp}) and the logarithmic bound we have in
(\ref{e:Jensenb}) becomes non-conclusive.  This gives a precise 
meaning for the statement at the end of the theorem 
and completes the proof.

\section{Derivation of the bounds for the PSP-approximation 
  of a Gaussian trace}\label{sec:canaccder}

In this section we first derive the ``saddle point approximation''
(\ref{e:PSPA}) for the Gaussian trace and then show how the two 
bounds for the accuracy of this approximation, 
(\ref{e:canJB}) and (\ref{e:Dcanbound}), can be obtained.  As 
in section \ref{sec:canacc}, let $L$ and $\lambda$ be
some parameters which define a Gaussian fluctuation spectrum.
Then $\beta_+=\infty$ and we shall inspect the ``canonical''
approximation of the Gaussian trace for any
$\beta>\beta_-$.  Let us also define $R=\lambda/\sigma(\beta)$.

By equation (\ref{e:trF}) we can then use the integral 
representation,
\begin{eqnarray}\label{e:trGl}
\fl \Tr\!\left[\A G_{\lambda}(L-g(\ham))\right] =
 \rme^{\beta L+ \half \lambda^2 \beta^2} 
 \int_{-\infty}^{\infty}
  \frac{\rmd \alpha}{ 2\pi} \rme^{- \half \lambda^2 \alpha^2} 
   \Tr\bigl(\A \rme^{-\beta g(\ham)} 
     \rme^{\ci\alpha(L+\beta\lambda^2-g(\ham))}\bigr).
\end{eqnarray}
The PSP-approximation is now
derived by expanding $\ln f(\beta+\ci\alpha)$ around
$\alpha=0$,  where $f(w) = \Tr\bigl(\A \rme^{-w g(\ham)}\bigr)$.  

$f(w)$ is an analytic function in the region $\re w > \beta_-$ 
whose restriction to the line $\im w=0$  is a strictly
positive function.  Although the logarithm of $f$ is possibly not
well-defined in the whole region, the following representation is
nevertheless valid for all $\beta>\beta_-$ and 
$\alpha\in\R$,
\begin{equation}\label{e:fspe}
f(\beta+\ci\alpha) = f(\beta) \rme^{-\ci\alpha \bar{g}(\beta)} 
 \exp\!\left[ -\alpha^2 
  \int_0^1\!\! \rmd t (1-t)  \sigma^2(\beta+\ci\alpha t)\right],
\end{equation}
where $-\bar{g}(w)$ and $\sigma^2(w)$, as defined by (\ref{e:defpars}), 
are the first and second derivative of $\ln f$.
If there is a singularity on the integration contour in
(\ref{e:fspe}), the integral has to be evaluated by an infinitesimal
deformation of the contour.   

Since $\sigma^2$ is analytic and $\sigma^2(\beta)> 0$, we can 
for all $0<r<1$ find a $\rho>0$ such that
\begin{equation}\label{e:defrc}
\left|\frac{\sigma^2(\beta+\ci\alpha)}{\sigma^2(\beta)}-1\right| 
 \le r \text{,\ for\ all\ } -\rho\le\alpha\le \rho.
\end{equation}
Let then $|\alpha|\le\rho$ and
define $z$ by $z= \half (r-1)\sigma^2(\beta)+ \int_0^1\!\! \rmd t
(1-t) \sigma^2(\beta+\ci\alpha t)$.  
By (\ref{e:defrc}), now $\re z \ge 0$
and $|z|\le r \sigma^2(\beta)$.  But since for every complex $w$ 
with $\re w\ge 0$ it is true that $|\rme^{-w}-1|\le |w|$, 
we have also the inequality
\[
\left|\rme^{-\alpha^2 z}-
  \rme^{- \smash{\half} r\alpha^2 \sigma^2(\beta)}\right| \le 
\left|\rme^{-\alpha^2 z}-1\right|+
 \left|\rme^{-\smash{\half} r\alpha^2 \sigma^2(\beta)}-1\right| \le
\frac{3}{2} r\alpha^2 \sigma^2(\beta).
\]
A comparison with (\ref{e:fspe}) then shows that we have proven the
following bound for the quadratic approximation of $f$,
\begin{equation}\label{e:qupbound}
\left| \frac{f(\beta+\ci\alpha)}{f(\beta) \rme^{-\ci\alpha
    \bar{g}(\beta)-\half\alpha^2 \sigma^2(\beta)}} - 1 \right| 
\le \frac{3}{2} r\alpha^2 \sigma^2(\beta)  
 \exp\left(\half r\alpha^2 \sigma^2(\beta)\right),
\end{equation}
valid for all $|\alpha|\le \rho$.
Since always $|f(\beta+\ci\alpha)|\le f(\beta)$,
it is also obvious that for all real $\alpha$ the left hand side is 
bounded by 
\begin{equation}\label{e:secondb}
\exp\left(\half \alpha^2 \sigma^2(\beta)\right) +1.
\end{equation}

Applying the quadratic expansion of $f$ to (\ref{e:trGl}) we get
\begin{eqnarray}\label{e:LSLS}
\fl \Tr\!\left[\A G_{\lambda}(L-g(\ham))\right] =
   \rme^{\beta L+ \half \lambda^2 \beta^2} 
 \Tr\!\left[\A\rme^{-\beta g(\ham)}\right] 
 G_{\sqrt{\lambda^2+\smash{\sigma^2(\beta)}}}
    (L+\beta\lambda^2-\bar{g}(\beta)) \nonumber\\
 \lo\times
  \left( 1 + \Delta_{\rm PSPA} \right), 
\end{eqnarray}
where $\Delta_{\rm PSPA}$ is defined by
\[
\fl
G_{\sqrt{\lambda^2+\smash{\sigma^2(\beta)}}}
  (L+\beta\lambda^2-\bar{g}(\beta))^{-1} 
 \int_{-\infty}^{\infty}
  \frac{\rmd \alpha}{ 2\pi} \rme^{- \half \lambda^2 \alpha^2 
     + \ci\alpha(L+\beta\lambda^2)}
 \biggl[ \frac{f(\beta+\ci\alpha)}{f(\beta)}- \rme^{-\ci\alpha
    \bar{g}(\beta)-\half\alpha^2 \sigma^2(\beta)}
  \biggr].
\]
By (\ref{e:qupbound}) and (\ref{e:secondb}) then
\begin{eqnarray}
\fl G_{\sqrt{\lambda^2+\smash{\sigma^2 }}}
  (L+\beta\lambda^2-\bar{g} ) |\Delta_{\rm PSPA}| \nonumber\\
 \lo\le \frac{3}{2} r \sigma^2  \int_{-\rho}^{\rho}
  \frac{\rmd \alpha}{ 2\pi} \alpha^2
     \rme^{- \half \alpha^2 (\lambda^2 +(1-r)\sigma^2  )}
  +  \int_{|\alpha|\ge\rho} \frac{\rmd \alpha}{ 2\pi}
     \biggl( \rme^{-\half\alpha^2\lambda^2} 
        + \rme^{-\half\alpha^2 (\lambda^2 +\sigma^2  )} 
     \biggr)  \nonumber\\
 \lo\le \frac{1}{\sqrt{2\pi}} \biggl[ \frac{3}{2} \frac{r}{\sigma} 
      \smash{\biggl(R^2+1-r\biggr)^{-\case32}} +
    \frac{\rme^{-\half\rho^2\lambda^2}}{\lambda}  +
    \frac{\rme^{-\half\rho^2(\lambda^2+\sigma^2)}
           }{\sqrt{\lambda^2+\sigma^2} } \biggr], \label{e:ineqs}
\end{eqnarray}
where we have extended the first integral over the whole real line and
used $\int \rmd x x^2 \exp[-\case12 x^2] = \sqrt{2 \pi}$ and
approximated the second integral as in
\[
 \int_{|\alpha|\ge\rho} \frac{\rmd \alpha}{ 2\pi}
     \rme^{-\half\alpha^2\lambda^2} = 
 2 \int_0^\infty \frac{\rmd y}{ 2\pi}      
   \rme^{-\half(y+\rho)^2\lambda^2} \le 
 \rme^{-\half\rho^2\lambda^2} 
  \int_{-\infty}^{\infty} \frac{\rmd y}{ 2\pi} 
    \rme^{-\half y^2\lambda^2}.
\]
The second bound (\ref{e:Dcanbound}) follows now easily from
(\ref{e:ineqs}).

The first bounds, inequalities
(\ref{e:canJB}), are a straightforward consequence of
the result
\begin{eqnarray*}
\fl
 \exp\biggl[
  -\frac{(L+\beta\lambda^2-\bar{g}(\beta))^2}{2(\lambda^2+\Lambda^2)} 
  -\frac{\sigma^2(\beta)}{2(\lambda^2+\Lambda^2)}\biggr]
 \le \frac{\Tr\bigl\{\A\exp\bigl[{-\frac{1}{2(\lambda^2+\Lambda^2)}
  (L- g(\ham))^2}\bigr]\bigr\}}{
   \rme^{\beta L +\half \beta^2 (\lambda^2+\Lambda^2) }  
 \Tr\bigl[\A\rme^{-\beta g(\ham)}\bigr]}
  \le  1,
\end{eqnarray*}
which was derived by using Jensen's inequality near equation
(\ref{e:lookforb}) in \ref{sec:gausspf}.

\section*{References}


\end{document}